\documentclass[conference]{IEEEtran}
\IEEEoverridecommandlockouts

\usepackage{cite}
\usepackage{microtype}
\usepackage{amsmath,amssymb,amsfonts}
\usepackage{algorithmic}
\usepackage[hyphens]{url}
\usepackage{graphicx}
\usepackage{textcomp}
\usepackage{xcolor}
\usepackage{nth}
\usepackage{subcaption} 
\usepackage{braket}
\usepackage{tabularx}
\usepackage{bm}
\usepackage{multirow}
\usepackage{makecell}
\def\BibTeX{{\rm B\kern-.05em{\sc i\kern-.025em b}\kern-.08em
    T\kern-.1667em\lower.7ex\hbox{E}\kern-.125emX}}
\begin{document}

\title{Distilling Magic States in the Bicycle Architecture
}

\author{
\IEEEauthorblockN{
Shifan Xu\IEEEauthorrefmark{1},
Kun Liu\IEEEauthorrefmark{2},
Patrick Rall\IEEEauthorrefmark{3},
Zhiyang He\IEEEauthorrefmark{4},
Yongshan Ding\IEEEauthorrefmark{1}\IEEEauthorrefmark{2}
}
\IEEEauthorblockA{\IEEEauthorrefmark{1}Department of Applied Physics, Yale University, New Haven, CT 06511, USA}
\IEEEauthorblockA{\IEEEauthorrefmark{2}Department of Computer Science, Yale University, New Haven, CT 06511, USA}
\IEEEauthorblockA{\IEEEauthorrefmark{3}IBM Quantum, IBM Research, Cambridge, MA 02142, USA}
\IEEEauthorblockA{\IEEEauthorrefmark{4}Department of Mathematics, Massachusetts Institute of Technology, Cambridge, MA 02139, USA}}

\maketitle

\begin{abstract}
Magic State Distillation is considered to be one of the promising methods for supplying the non-Clifford resources required to achieve universal fault tolerance. Conventional MSD protocols implemented in surface codes often require multiple code blocks and lattice surgery rounds, resulting in substantial qubit overhead, especially at low target error rates.

In this work, we present practical magic state distillation factories on Bivariate Bicycle (BB) codes that execute Pauli-measurement-based Clifford circuits inside a single BB code block. We formulate distillation circuit design as a joint optimization of logical qubit mapping, gate scheduling, measurement nativization, and protocol compression via qubit recycling. 
Based on detailed resource analysis and simulations, our BB factories have space-time volume comparable to that of leading distillation factories while delivering lower target error at a smaller qubit footprint, and are particularly compelling as second-round distillers following magic state cultivations.
\end{abstract}


\section{Introduction}
Building large-scale quantum computers, which can realize transformative applications such as factoring~\cite{shor1994algorithms}, requires the use of quantum error correction to suppress physical noise and enable universal, fault-tolerant quantum computation (FTQC)~\cite{shor1996fault-tolerant,kitaev1997quantum2}. 
To protect quantum information from decoherence, foundational works~\cite{shor1995scheme,steane1996multiple-particle,calderbank1996good,gottesman1997stabilizer} showed that we can encode information in \textit{stabilizer codes}, and perform repeated quantum measurements and classical decoding to correct from arbitrary physical errors. 
These schemes enabled the development of threshold theorems for FTQC~\cite{aharonov1997fault-tolerant}, which states that arbitrarily large quantum computation can be realized through QEC, assuming that physical error rates are below a constant threshold.
Among the many codes developed, the surface code~\cite{bravyi1998quantum,dennis2002topological,eczoo_surface} has been at the center of QEC research for the past two decades due to its promising practical performance, including low connectivity requirement, high threshold and fast decoding algorithms. 
Despite these advantages, surface code incurs a significant space overhead in realizing FTQC: factoring 2048-bit integers in surface code architectures uses physical qubits on the scale of millions~\cite{gidney2021how,gidney2025factor}.
More recently, significant research has studied quantum low-density parity-check (LDPC) codes, which can realize FTQC with low space overhead~\cite{gottesman2014fault-tolerant}. 
This asymptotic promise led to the development of many families of QLDPC codes with practical parameters~\cite{eczoo_qldpc}, notably the Bivariate Bicycle (BB) codes introduced by IBM~\cite{bravyi2024high-threshold2}.

To perform computation on encoded quantum information, many schemes and architectures have been proposed for surface codes~\cite{dennis2002topological,fowler2012surface,fowler2018low,litinski2019game,zhou2024algorithmic} and QLDPC codes~\cite{gottesman2014fault-tolerant,fawzi2018constant,tamiya2024polylog-time-and,nguyen2024good,he2025composable,zhang2025accelerating,he2025extractorsqldpcarchitecturesefficient,yoder2025tour}. 
An essential component common to almost all existing architectures is the production of high-fidelity non-Clifford resource states known as \emph{magic states}.
These magic states can be consumed through gate teleportation~\cite{gottesman1999demonstrating} to fault-tolerantly implement non-Clifford logical gates on encoded information.
Conceptually, gate teleporting magic states is used in most architectures because non-Clifford gates, without which we cannot perform universal quantum computation, are at this time more costly to implement using other common mechanisms such as transversal gates. 
Gate teleportation serves as an approach where most of the cost of non-Clifford gates is offloaded to magic state preparation, while the active cost at runtime (namely performing the teleportation) is relatively lower.  
In most architectures, the total overhead and logical gate speed are often bottlenecked by the costs of the magic state factories. 

The most widely applied method of magic state production is called \textit{magic state distillation} (MSD)~\cite{bravyi2005universal}, which consumes many copies of noisy, lower-fidelity magic states to produce fewer copies of higher-fidelity magic states. 
This procedure is applied to logical qubits, and typically iterated to suppress logical error rate to the scale needed for large scale FTQC ($\le 10^{-12}$). 
Conventional MSD factories are therefore very resource-intensive, often becoming the architectural bottleneck.
An alternative approach called \textit{magic state cultivation} (MSC)~\cite{gidney2024magicstatecultivationgrowing} gained recent attention due to its impressive practical performance, achieving a logical error rate of $2\times 10^{-9}$ at $10^{-3}$ physical error rate. 
MSC injects a physical magic state into a surface code logical qubit, and suppresses infidelity by applying multiple rounds of post-selection and physical non-Clifford gates.
Due to its use of exponentially scaling post-selection, MSC does not scale asymptotically and may not suffice for large scale FTQC.
\begin{figure}[t]
    \centering
    \includegraphics[width=\linewidth]{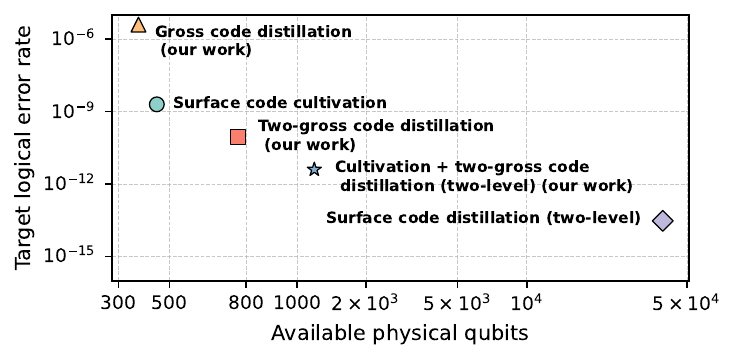} 
    \caption{Magic-state factory design space. Target logical error rate as a function of available physical qubits for surface-code cultivation, BB-code distillation on gross and two-gross codes, and two-level protocols combining cultivation with two-gross or surface-code-only distillation at physical error rate $p_{\mathrm{phys}}=10^{-3}$. Our BB-based factories achieve lower output error at similar qubit budgets, and the cultivation + two-gross pipeline extends to lower error regimes than cultivation can reach.}
    \label{fig:motivation}
\end{figure}

In this work, we introduce new designs of magic state factories that integrate MSC on surface code and MSD in BB codes, achieving low logical error rates while using significantly fewer resources compared to conventional distillation factories.
In more detail:
\begin{itemize}
    \item We present a collection of MSD protocols that run entirely within a single BB code block\footnote{More specifically, the MSD circuits are supported within a single BB code block, while the input magic state may be supported externally.}, in contrast to conventional factories which run MSD on multiple blocks of QLDPC codes or surface codes.
    This one block design confers significant savings in the space overhead of magic state factories, reducing the physical qubit count from thousands to hundreds. Additionally, it simplifies the architecture, decreases the number of long range inter-block connections, and reduces decoding complexity.

    \item We present comprehensive, end-to-end optimizations for existing MSD circuits, including (i) logical qubit mapping to maximize the use of native, low cost logical gates\footnote{These gates are performed through code surgery with the help of an ancilla system, which we detail in Section~\ref{sec:background}.} on the BB code, 
    (ii) a masking technique that augments rotations with $Z$ on inactive qubits to turn a set of sparse native logical operation on all logical qubits into a denser set over the active logical qubits, 
    (iii) gate scheduling cast as a Traveling Salesman Problem to minimize compiled circuit depth, 
    and (iv) parallel execution of two MSD protocols on the same BB code block.

    Together, these methods significantly reduce the circuit depth of MSD and improve the factory throughput.

    \item We introduce a general method to compress MSD circuits to be supported on fewer logical qubits while maintaining the same level of logical error suppression. 
    As examples, we compress the 49-to-1 protocol from $13$ to $7$ qubits, the 51-to-3CS protocol from $18$ to $9$ qubits,the 64-to-2CCZ protocol from $17$ to $10$ qubits, making these circuits implementable within a single BB code block. Moreover, this method is generic to any magic state distillation protocol generated by tri-orthogonal matrices.
    As near-term quantum computers are space-limited, our method brings large MSD protocols significantly closer to practice.

    \item We benchmark a collection of magic state factories with varying protocols, codes, and noise levels. Our results quantify the substantial overhead reduction enabled by the proposed optimizations, and, through sensitivity analysis, provide practical guidance for protocol design under different operating regimes, while highlighting the key bottlenecks and improvement opportunities for each design.
\end{itemize}

As shown in Figure~\ref{fig:motivation}, our leading proposal is to perform 15-to-1 MSD on a block of two-gross code (an instance of BB code), using logical magic states supplied by MSC on a surface code. 
Under $10^{-3}$ physical error rate, this protocol is estimated to reach logical error rates lower than what is currently achievable through MSC, while using just one block of surface code and one block of two-gross code.
Our factories naturally fit into the bicycle architecture~\cite{yoder2025tour}, a promising near-term FTQC architecture based on the BB codes~\cite{bravyi2024high-threshold2}. 
Conceptually, the bicycle architecture encodes information in BB codes and performs logical operations using recently-developed code surgery techniques~\cite{cross2024improvedqldpcsurgerylogical,williamson2024low-overhead,ide2024fault,swaroop2025universaladaptersquantumldpc}. 
These operations consume magic states, which makes magic generation the central component that determines the overall efficiency. Our designs thereby constitute an essential improvement to the bicycle architecture. For general extractor architectures~\cite{he2025extractorsqldpcarchitecturesefficient}, our techniques can also be applied to design other efficient magic state factories.

The rest of this paper is organized as follows. Sec.~\ref{sec:background} reviews the fundamentals of quantum error correction, the BB code family, and magic state distillation. In Sec.~\ref{sec:MSD}, we present a full stack BB code based magic state distillation factory design. Sec.~\ref{sec:optimization} introduces multiple optimization methods to reduce both the depth and the size of the MSD circuit. Sec.~\ref{sec:compression} introduces a protocol-level compression method to reduce the logical-qubit footprint of the MSD circuit by recycling inactive logical qubits. In Sec.~\ref{sec:eval} and Sec.~\ref{sec:results}, we evaluate the space time cost and output logical error rates of our approach and compare them with surface code based magic state distillation and cultivation.

\section{Background}\label{sec:background}

\begin{figure}[t]
         \centering
         \includegraphics[width=0.485\textwidth]{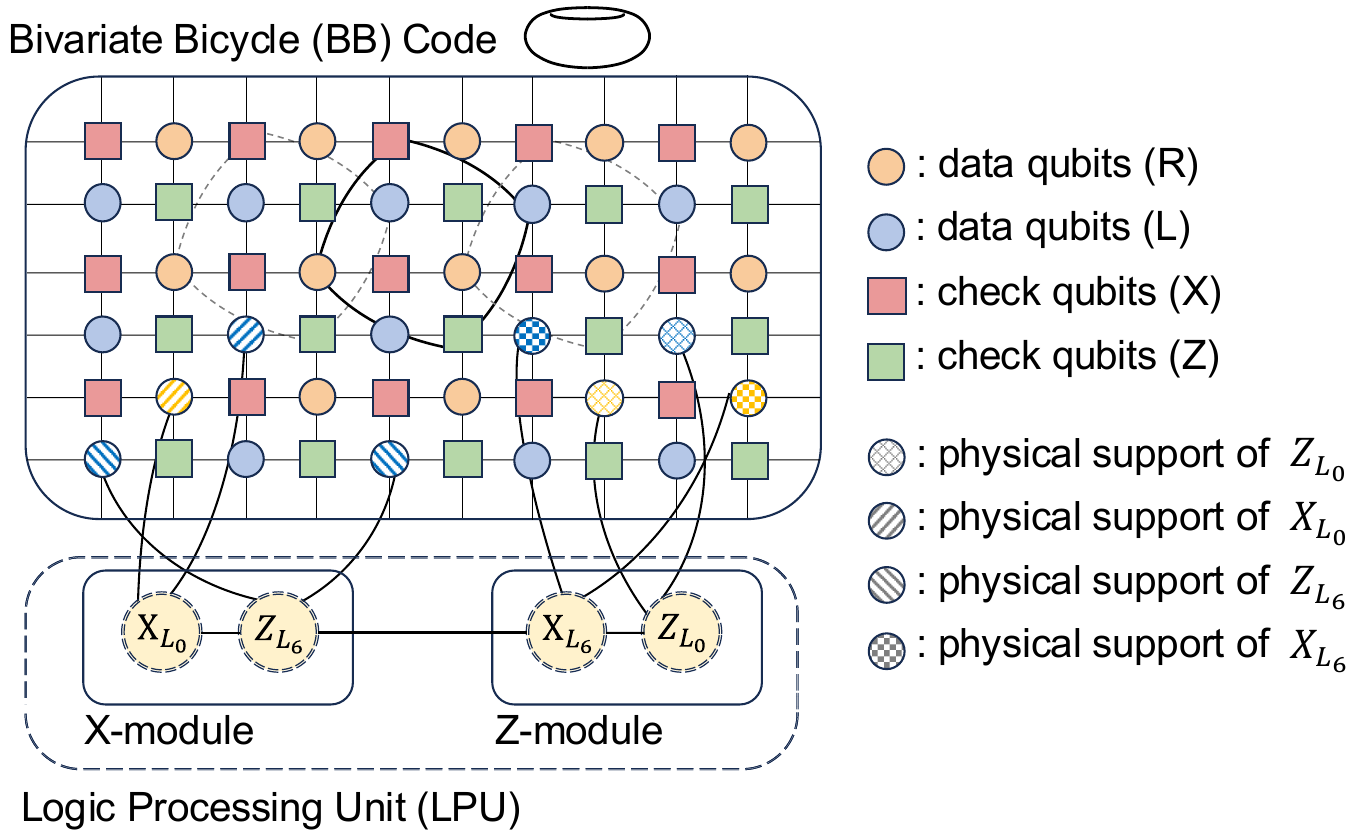}
             \caption{
             One module in the bicycle architecture, consisting of a Bivariate Bicycle (BB) code and a Logic Processing Unit (LPU). 
             BB codes can be presented \cite{bravyi2024high-threshold2} as a lattice of physical qubits on a torus, consisting of X and Z check qubits and L and R data qubits. Connections between qubits are within the lattice and also along certain long-range connections.
             The LPU, which consists of two modules, is connected to the $X$ and $Z$ operators of the pivot qubit $L_0$ and its dual $L_6$. 
             The LPU is capable of measuring any product of operators $X_{L_0}, Z_{L_0}, X_{L_6},$ and $ Z_{L_6}$.
             See Section~\ref{subsec:cliffordsynth} for more details.
             } 
    \label{fig:bg}
\end{figure}

\subsection{Quantum Error Correction}
Quantum Error Correction (QEC) is crucial for the development of scalable quantum computation \cite{shor1996fault-tolerant}. A quantum error correction code has parameters $[[n, k, d]]$, where $n$ is the number of physical qubits in hardware, $k$ is the number of encoded logical qubits which experience suppressed error rates and are used for computation, and the distance $d$ is a measure for the degree of error suppression. 
The standard surface code has parameters $[[d^2+(d-1)^2, 1, d]]$ for variable $d$.

Quantum low-density parity-check (QLDPC) codes, generalized from classical low-density parity-check (LDPC) codes, are particularly attractive for scalable fault-tolerant quantum computing due to their high encoding rates, which reduce the space overhead of FTQC, and sparse check structure, which allows for efficient decoding algorithms and locality in physical implementation. 
Recent advances in constructing high-rate codes~\cite{panteleev2021degenerate,bravyi2024high-threshold2,xu2024constant-overhead,lin2024quantum}, implementing logical operations~\cite{Breuckmann_2024,williamson2024low-overhead,he2025extractorsqldpcarchitecturesefficient,xu2025batchedhighratelogicaloperations}, and building hardware with long-range connections~\cite{bravyi2022future,psiquantum2025manufacturable,xanadu-aghaee2025scaling,Bluvstein_2023} have established QLDPC codes as promising candidates for near-term quantum memories and FTQC.

\subsection{Bivariate Bicycle Codes}
\label{subsec:bivariate-bicycle}

The Bivariate Bicycle (BB) codes are a family of recently proposed QLDPC codes \cite{bravyi2024high-threshold2}, with the ``gross'' [[144,12,12]] and ``two-gross'' [[288,12,18]] codes being particularly attractive due to their high encoding rate and promising hardware realizability. 
Schemes for implementing logical operations vary significantly across different code families, so we focus our attention on BB codes for this work.

We give a brief mathematical definition of BB codes. Consider an abelian group $\mathcal{M} = \{x^i y^j \mid i \in [0,\ell-1],\, j \in [0,m-1]\}$, where indices are taken modulo $\ell$ and $m$. 
This group admits a representation in terms of permutation matrices $\mathbb{F}_2^{\ell m\times \ell m}$. 
To specify a BB code, we select $\ell, m$ and elements $A_1, A_2, A_3, B_1, B_2, B_3$ of $\mathcal{M}$. 
We form the matrices $A = A_1 + A_2 + A_3$ and $B = B_1 + B_2 + B_3$ via the permutation representation, and define parity-check matrices as $H_X = [A \mid B]$ and $H_Z = [B^\top \mid A^\top]$, where $\top$ denotes transposition. The $n = 2\ell m$ columns of these matrices denote physical qubits, and the rows denote supports of $X$- and $Z$-type stabilizers for $H_X$ and $H_Z$ respectively. 
By construction, all checks are supported on six data qubits and all data qubits participate in six checks. This leads to a natural realization on superconducting architectures with $2\ell m$ data qubits and $2\ell m$ check qubits, which are the vertices of a connectivity graph with degree six.

\subsection{Fault-tolerant Architectures Based on QLDPC Codes}
\label{subsec:cliffordsynth}

Fault-tolerant computation on QLDPC codes can be realized through fault-tolerant measurement of logical Pauli operators assisted with high-fidelity magic states~\cite{bravyi2016trading}. 
To implement logical measurements, a popular technique is generalized code surgery~\cite{Cohen_2022, cross2024improvedqldpcsurgerylogical, williamson2024low-overhead,ide2024fault,swaroop2025universaladaptersquantumldpc,  he2025extractorsqldpcarchitecturesefficient, yoder2025tour}, which appends a carefully constructed ancillary system of physical qubits to act as a measurement gadget. 
Recent developments in surgery have introduced techniques to construct highly flexible ancilla systems, which led to the proposal of the bicycle architecture~\cite{yoder2025tour} and its generalization, extractor architectures~\cite{he2025extractorsqldpcarchitecturesefficient}.
In these architectures, logical information is encoded in multiple QLDPC code blocks, each augmented by an ancilla system called an extractor or, for BB codes, a Logic Processing Unit (LPU). These ancilla systems are connected together to enable flexible measurements of many logical operators and, for universal FTQC, are further connected to magic state factories. 
Our work focuses on the bicycle architecture and its magic state factories.

The bicycle architecture \cite{yoder2025tour} relies on a combination of a highly optimized LPU with fault-tolerantly realizable unitary automorphism gates. The LPU construction exploits symmetries in the Bivariate Bicycle codes to maximize measurement capabilities while minimizing additional physical qubit count. For example, the [[144,12,12]] code admits an LPU with 90 qubits, and the [[288,12,18]] code admits an LPU with 158. We briefly discuss the logical gate set in the bicycle architecture.

\subsubsection{Automorphism gates}
An automorphism gate is a permutation of physical qubits which enacts a \texttt{CNOT} circuit on the logical qubits. Within the group of automorphisms, we focus on a 36-element subgroup of \emph{shift automorphisms} that preserve $X$/$Z$ type. 
Twelve of these can be implemented fault-tolerantly using the same sparse connectivity as syndrome extraction. These 12 elements form a convenient generating set: any shift automorphism can be synthesized using at most two generators.

A key structural feature of BB codes is their ZX-duality \cite{Breuckmann_2024}: an order-two permutation of physical qubits followed by a layer of Hadamards maps performs a logical gate. A basis for the twelve logical qubits is chosen so that they divide into two blocks of six, and these two blocks are swapped and Hadamarded by the ZX-duality. Since the shift automorphisms commute with the ZX-duality, we know their logical unitaries take the form $U_{\mathrm{aut}} \otimes U_{\mathrm{aut}}$ on the two logical blocks.

\subsubsection{Native measurements and rotations}
The LPU enables $15$ different Pauli measurements supported on two logical qubits, see Figure~\ref{fig:bg}. Composing these measurements with the $36$ shift automorphisms gives us $540 = 15\times 36$ native multi-qubit Pauli measurements on up to 12 logical qubits.
Although this is only a small subset of the full $4^{12}$ Pauli group, they generate all Clifford operations on 11 logical qubits through a measurement-to-rotation compilation.
As shown in \cite{cross2024improvedqldpcsurgerylogical} and Fig.~\ref{fig:meas_to_rotation}(d), using one logical qubit as a \textit{pivot qubit}, we can implement $\exp(i \frac{\pi}{4} P)$ rotations for multi-qubit Pauli $P$ by measuring $P$ on the desired qubits and $X$, $Y$, $Z$ on the pivot qubit. 
These \textit{native rotations} implemented by \textit{native measurements} can then be composed to generate the Clifford group on 11 logical qubits.

\begin{figure*}[t]
    \centering
    \includegraphics[width=\textwidth]{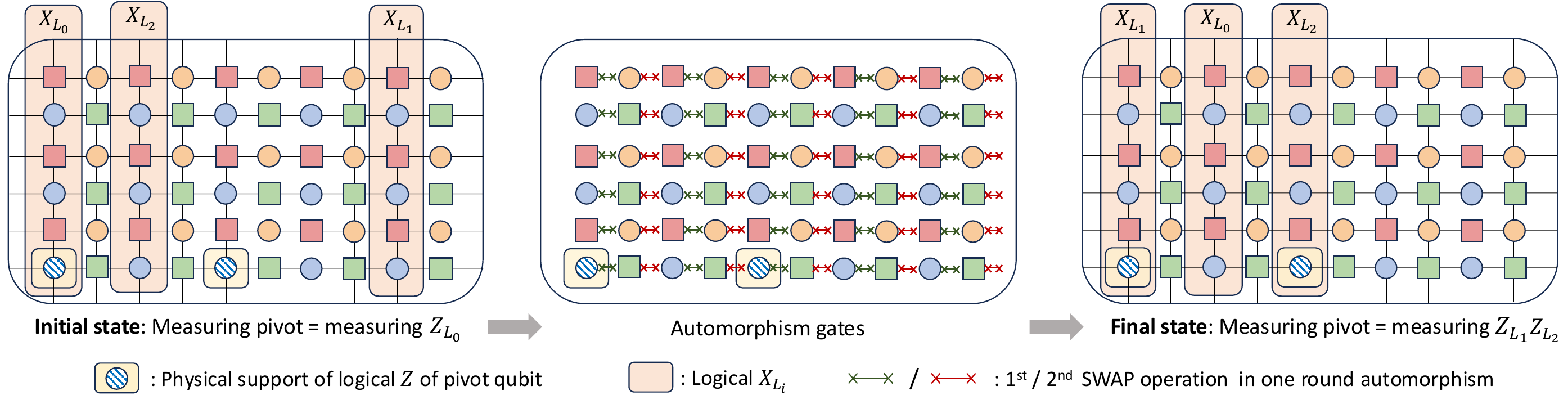}
    \caption{
    Fault-tolerant implementation of a shift-automorphism generator and its impact on logical operators. Shift automorphisms permute data qubits via successive swap operations (green, then red) between data and check qubits along edges in the connectivity graph. Logical operators $X_{L_0}, X_{L_1}, X_{L_2}$ supported on shaded regions are permuted so that their overlap with the pivot's $Z_{L_0}$ support changes. After conjugation, multi-qubit Paulis that were not directly accessible through the pivot become measurable via an LPU $Z_{L_0}$ measurement.}
    \label{fig:automorphism}
\end{figure*}

\subsubsection{Simultaneous measurements}
\label{sec:simultaneous-measure}

The LPU attaches independently to two logical qubits, the pivot and its dual, and consists of separate $X$ and $Z$ modules. This naturally suggests that certain pairs of Pauli operators may be measured simultaneously. For example, an $X \otimes X$ measurement across the pivot and its dual can be realized by coupling the $X$-module and $Z$-module through a transient bridge system. Removing the bridge implements two single-qubit measurements, $X \otimes I$ and $I \otimes X$, in one logical cycle. Analogous constructions exist for $Z \otimes I$ and $I \otimes Z$.
We note that simultaneous measurements can reduce circuit depth and amortize LPU cost.

The native gate sets in the bicycle architecture are benchmarked in~\cite{yoder2025tour}. 
Simulation results indicate that automorphisms are substantially less error-prone than LPU measurements, although their cost depends on the number of generators used.
Automorphism gates are also much faster than LPU measurements.
Our optimization and compilation of MSD protocols therefore aim to minimize the number of LPU measurements. 
\subsection{Magic State Distillation}

In almost all modern FTQC architectures, logical non-Clifford gates, such as the $T=\mathrm{diag}(1,e^{i\pi/4})$ gate, are implemented through various forms of gate teleportation~\cite{gottesman1999demonstrating}. 
These protocols consume magic states, such as $\ket{T}=(\ket{0}+e^{i\pi/4}\ket{1})/\sqrt{2}$ for the $T$ gate.
Efficient preparation of high-fidelity magic states is therefore crucial for the performance of these architectures. 
For most quantum codes, preparation of high-fidelity magic states is not possible directly in an error-corrected setting.
However, noisy magic states can be prepared outside of the protection of an error-correction code, and then loaded into a logical qubit so further errors do not accumulate. 
Such noisy state injection produces $\ket{T}$ with error rate $p_{\mathrm{inj}}$ that is typically much larger than the target logical error rate $p_{\mathrm{L}}$ required for an algorithm. 
To bridge this gap, one uses \emph{magic state distillation} (MSD): a fault-tolerant protocol that consumes multiple noisy copies of a magic state to produce fewer, higher-fidelity outputs. Intuitively, MSD trades quantity for quality, suppressing errors by leveraging redundancy and special algebraic structure in the underlying code.

Most magic state distillation protocols are constructed from $[[n,k,d]]$ stabilizer codes, which are sometimes called distillation codes. 
Unlike codes that are used for quantum memory like BB codes, distillation codes feature additional algebraic constraints such as featuring a transversal $T$ gate, allowing them to map $n$ noisy input magic states to $k$ improved output magic states. A single round achieves output error of the form $p_{\mathrm{out}} \approx c\,p_{\mathrm{inj}}^t + O(p_{\mathrm{L}})$, 
where $t = O(d)$ is the degree of error suppression (in many protocols $t = d$), $c$ is a protocol-dependent constant, and $p_{\mathrm{L}}$ captures additional faults from running the MSD circuits on logical qubits. A notable $[[15,1,3]]$ protocol distills $\ket{T}$ states with $c=35$ and $t = 3$~\cite{bravyi2005universal}.

Repeated application of a distillation protocol results in substantial error suppression. Chaining $r$ rounds yields $p_{\mathrm{out}}^{(r)} \sim \tilde{c}\,p_{\mathrm{inj}}^{t^r}$ at the cost of increased qubit and time overhead. For this reason, up to two rounds of distillation suffice for many applications.

\section{Overview: Magic State Distillation in BB Architectures}
\label{sec:MSD}
We design magic state distillation factories tailored to the Bivariate Bicycle (BB) architectures of \cite{cross2024improvedqldpcsurgerylogical, yoder2025tour}. Our goal is high-fidelity, low-latency $\ket{T}$-state distillation that respects the locality, modular structure, error pathways, and syndrome-extraction capabilities of BB codes. We first show how standard MSD protocols map cleanly onto the BB layout, then present several techniques for implementing the required multi-qubit $\exp(i \frac{\pi}{8} P)$ rotations via injected $\ket{T}$ states. Throughout, we highlight how inter-module connectivity, pivot-qubit access, and measurement fidelity influence factory design.

\subsection{Magic State Distillation with Triorthogonal Matrices}

We focus on $\ket{T}$-state distillation following the construction of \cite{Litinski_2019}, where each protocol is defined by a \emph{triorthogonal} matrix \cite{Bravyi_2012}. 
Let $G \in \{0,1\}^{m \times n}$ be such a matrix with $k$ rows of odd weight (corresponding to the $k$ output qubits) and $m{-}k$ rows of even weight (corresponding to ancilla qubits).
Each column $c$ specifies a commuting $Z$-type $\pi/8$ rotation acting on the set of qubits/rows $S_c=\{r : G_{rc}=1\}$.

Given $G$, the protocol proceeds as follows:

\begin{enumerate}
    \item Initialize $m$ logical qubits to $\ket{+}$. In the BB architecture, this is achieved by preparing all physical qubits in $\ket{+}$ and running one syndrome-extraction cycle.

    \item For each column $c$, consume one noisy $\ket{T}$ and implement $e^{i \frac{\pi}{8} P}$.
    Here $P = Z^{\otimes S_c}$ is an $m$-qubit Pauli with $Z$ on rows in $S_c$ and identity elsewhere. 
    \item Measure the $m{-}k$ parity check qubits in the $X$ basis and postselect on the all-$\ket{+}$ outcome. When postselection succeeds, the first $k$ rows contain the distilled $\ket{T}^{\otimes k}$. 
    In BB codes, these measurements can be implemented in $k{+}1$ logical steps as follows: for each output magic state, fix a target qubit (on another code block) and measure $Z\!\otimes\!Z$ between the magic state and the target qubit; then measure all physical qubits in the $X$ basis. This procedure projects the output magic states onto their destinations.
\end{enumerate}

This formulation uses only $m$ logical qubits, unlike the original method of \cite{Bravyi_2012}, which prepared an $n$-qubit stabilizer state, applied $T$ to each qubit, and unencoded via Clifford operations. It preserves the same distillation performance with far fewer qubits and Clifford gates, making it well-suited to BB codes where logical qubit count is tightly constrained.

\begin{figure*}[t]
    \centering
    \includegraphics[width=0.9\textwidth]{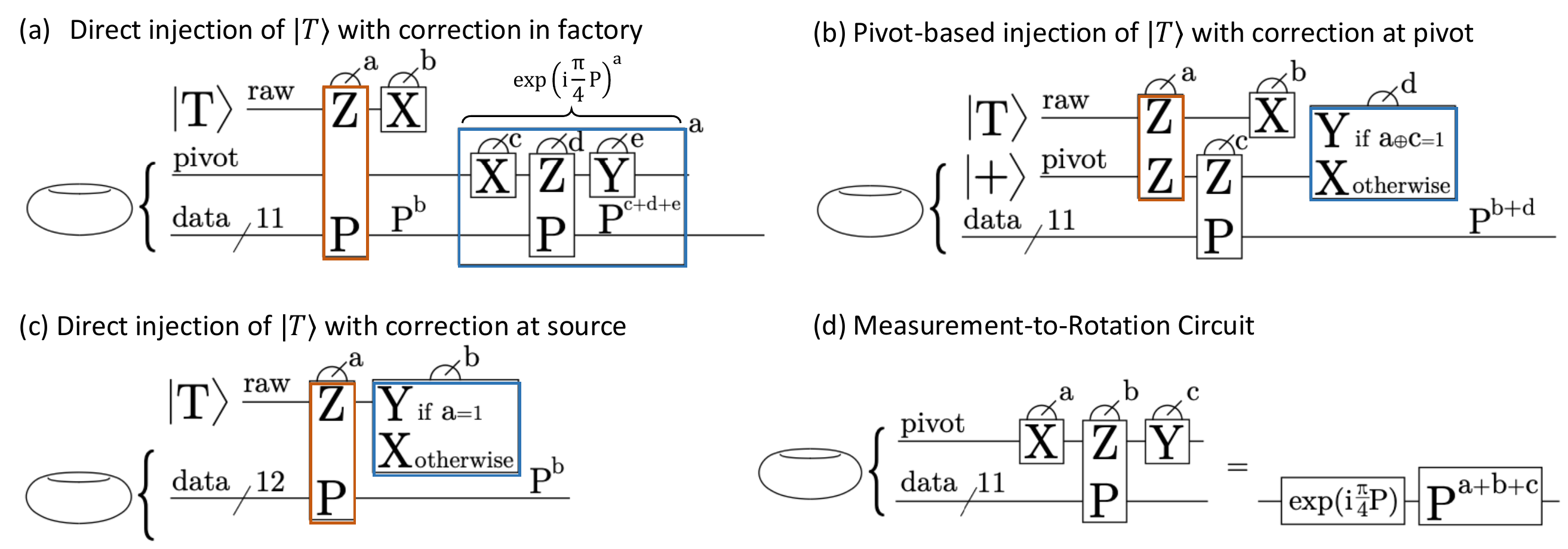}
    \caption{
    (a–c) Magic state injection schemes for implementing $\exp(i \frac{\pi}{8} P)$ in a BB architecture. The schemes differ in how the magic state is teleported to the target qubits and how the resulting conditional Clifford correction is handled, leading to different latency and error profiles. Inter-module measurements are shown in orange, conditional Clifford corrections in blue. A $P$ label denotes a Pauli operator, which is cheap to track in fault-tolerant architectures. (d) Measurement-to-rotation circuit implementing $\exp(i \frac{\pi}{4} P)$ using a designated pivot qubit and BB's toric symmetry. We use pivot injection as the default throughout the paper because it achieves a lower error rate, as demonstrated by the detailed benchmarking results in Sections~\ref{sec:factory-error-analysis} and \ref{sec:results}.}
    \label{fig:magic_inject}
    \label{fig:meas_to_rotation}
\end{figure*}

\subsection{Implementing $\pi/8$ Rotations}

Implementing the protocol reduces to realizing $\exp(i \frac{\pi}{8} P)$, where $P = Z^{\otimes S_c}$. We adapt three approaches from \cite{Litinski_2019} to the BB architecture (Figure~\ref{fig:magic_inject}). The key distinction among them is how they handle the necessary conditional Clifford correction: standard injection produces either $\exp(+i \frac{\pi}{8} P)$ or $\exp(-i \frac{\pi}{8} P)$ at random, so a corrective $\exp(i \frac{\pi}{4} P)$ may be required. The three approaches are as follows:

\begin{itemize}
    \item \textbf{Direct injection with factory correction: Fig.\ref{fig:magic_inject}(a).}  
    The input $\ket{T}$ is teleported directly onto all target qubits via an inter-module measurement. Any required correction is implemented explicitly using the measurement-to-rotation circuit (Fig.~\ref{fig:meas_to_rotation}(d)). The injection itself does not require the pivot, but the correction does.

    \item \textbf{Pivot-based injection with pivot correction: Fig.\ref{fig:magic_inject}(b).}  
    The $\ket{T}$ state is first teleported onto the pivot qubit, then onto the target qubits using only in-module measurements. This confines the noisy injection step to a single qubit and avoids spreading error across the data block. The correction is absorbed into a conditional $X$ or $Y$ measurement on the pivot. 

    \item \textbf{Direct injection with source correction: Fig.\ref{fig:magic_inject}(c).}  
    If the module supplying the $\ket{T}$ states supports direct, high-fidelity $Y$ measurements, the correction can be applied by the source qubit itself. In this case, the pivot is not involved at all, and no additional correction step is required. Whether this is viable depends on the native measurement bases of the $\ket{T}$-state source.
\end{itemize}

These three strategies span different hardware assumptions and error models. In Section~\ref{sec:factory-error-analysis}, we quantify how their measurement counts, routing demands, and error locations translate into overall factory throughput and logical error rates for realistic BB-code parameters. In Section~\ref{sec:results}, we present detailed benchmarks showing how different injection schemes can be selected adaptively under different hardware assumptions, and we explore the tradeoff between spacetime volume and output error rate across these schemes.

\section{Implementation and Optimizations}
\label{sec:optimization}

In this section, we present techniques that improve the efficiency and reliability of magic state distillation within the bicycle architectures. These optimizations address bottlenecks from restricted native measurements, limited logical qubits, and architectural error sources. Together, they define a practical workflow for compiling distillation protocols into fault-tolerant BB-code factories with minimal overhead.

\subsection{Logical Qubit Mapping: Maximizing Native Coverage}
\label{subsec:logicalmapping}
Because the native measurement set is limited, not every Pauli $P$ required by the protocol can be realized by a single LPU measurement, even after conjugation. However, the protocol uses only a fixed set of logical qubits, which we are free to place within the BB code.

We therefore treat logical-qubit placement as an optimization problem. Given a BB code with $k$ data qubits and an $m$-qubit distillation protocol, we choose an $m$-element subset $S \subseteq [k]$ and assign protocol qubits to $S$ so that the number of native Pauli rotations is maximized. For the small protocol sizes of interest, we can brute-force over $S$; ties are broken by minimizing routing distance to the pivot.

For example, in the 15-to-1 protocol on the gross code, choosing 5 of the 6 qubits in a logical block yields native realizations for most of the required $\pi/8$ rotations, with the remainder implemented either by Clifford conjugation \cite{cross2024improvedqldpcsurgerylogical} or by masking (Section~\ref{subsec:masking}). This mapping step is lightweight but important: improving native coverage directly reduces factory latency and logical error.

\subsection{Masking: Enabling More Native Measurements}
\label{subsec:masking}

When $m < k$, unused logical qubits can be repurposed to expand the effective native measurement set. Suppose a required rotation $\exp(i \frac{\pi}{8} P)$ is non-native, but there exists a native Pauli $Q$ that matches $P$ on the $m$ active qubits and differs only on a subset of idle qubits. For instance, if those idle qubits are initialized to $\ket{0}$, then applying $Z$ on them leaves the state invariant, since $Z\ket{0} = \ket{0}$. We can therefore replace $P$ by $Q = P \cdot \prod_{j \in \mathcal{M}} Z_j$,
where $\mathcal{M}$ is a set of masked qubits chosen so that $Q$ is native.

This \emph{masking} operation is purely logical and adds no depth. It increases the fraction of rotations that can be executed as single native measurements.

\begin{figure}[t]
    \centering
    \includegraphics[width=\linewidth]{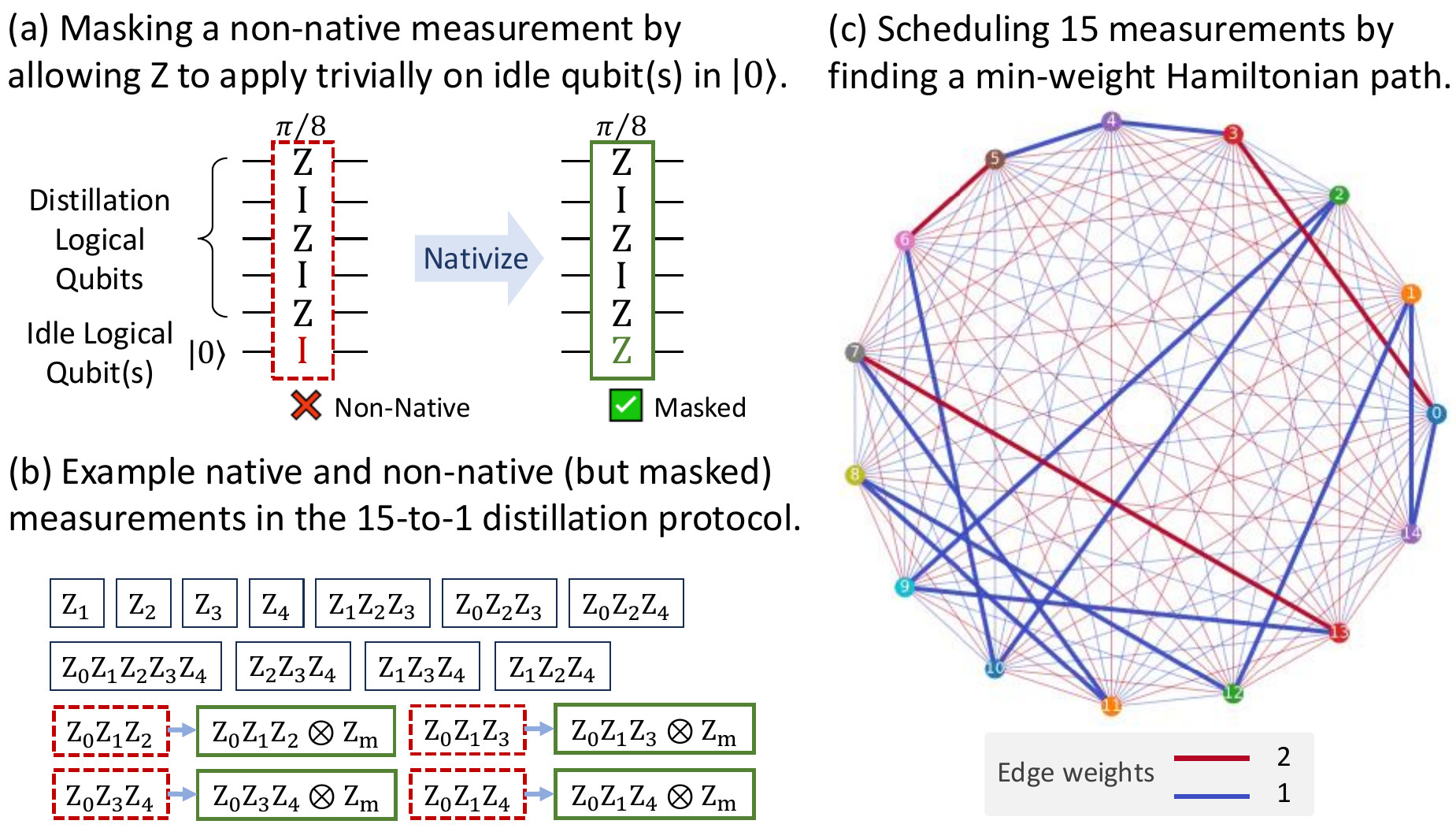}
    \caption{
    (a) Masking technique to nativize a Pauli measurement by allowing $Z$ to act on an idle logical qubit initialized to $\ket{0}$ within the BB code. (b) Native and non-native measurements in the 15-to-1 distillation circuit, which becomes fully nativized after masking. (c) Scheduling the 15-to-1 rotations in an order that minimizes automorphism rounds between successive measurements.}
    \label{fig:tspmask}
\end{figure}

In the 15-to-1 protocol, masking fully nativizes all 15 rotations: the four previously non-native Paulis become native when augmented with $Z$ factors on masked qubits.  As illustrated in Figure~\ref{fig:tspmask}, masking allows each rotation to be implemented using a single automorphism sequence and one LPU measurement (up to tracked byproduct Paulis), eliminating the need for additional Clifford conjugation.

\subsection{Gate Scheduling: Reducing Automorphism Rounds}
\label{subsec:tsp}

As above, a measurement of a logical Pauli $P$ is implemented by conjugating an LPU-native measurement with one or more automorphism gates. Different automorphisms incur different costs, typically corresponding to one or two automorphism-generator applications. In injection schemes that do not require intermediate pivot measurements between successive $\exp(i \frac{\pi}{8} P)$ gates, such as direct injections (Figure~\ref{fig:magic_inject}), we can reduce total automorphism cost by optimizing the order of Pauli rotations.

All $\exp(i \frac{\pi}{8} P)$ gates in a triorthogonal distillation protocol commute, so we are free to reorder them without changing the logical channel. The scheduling problem thus reduces to finding an execution order that minimizes the cumulative automorphism overhead needed to retarget the LPU between consecutive measurements.

We model this as a graph problem. Each distinct Pauli label $P$ in the protocol is represented as a node $v$ in a directed graph $G = (V,E)$. For any ordered pair $(u,v)$, we define the edge weight $w(u,v)$ as the cost of transforming the measurement configuration for $u$ into that for $v$ using automorphisms. This cost can be defined in terms of the number of automorphism rounds, latency, or any hardware-informed metric.

Any ordering of the rotations corresponds to a permutation $\sigma$ of the nodes in $V$, with total routing cost
\[
C(\sigma) = \sum_{i=1}^{|V|-1} w\big(v_{\sigma(i)},v_{\sigma(i+1)}\big).
\]
Minimizing $C(\sigma)$ is equivalent to finding a minimum-cost Hamiltonian path from $v_{\sigma(1)}$ to $v_{\sigma(|V|)}$, that is, a Traveling Salesman Problem (TSP) instance with fixed endpoints.

The TSP formulation changes only the measurement order, not the gates themselves or their angles. Although TSP is NP-hard in general, our instances are small; for example, the 15-to-1 protocol has only fifteen distinct $\exp(i \frac{\pi}{8} P)$ rotations. For such sizes, standard heuristics such as nearest-neighbor initialization with 2-opt or 3-opt refinements, or a warm-started mixed-integer linear program, quickly find near-optimal or optimal routes. The automorphism cost matrix can be precomputed once per BB-code instance and reused across factory cycles, so the marginal scheduling overhead is negligible.

\subsection{Improving Throughput: Multi-Track Distillations}
\label{subsec:multi-track}

When the triorthogonal matrix $G$ has small row count $m$, the BB architecture can host multiple protocol instances in parallel on a single code block. For the 15-to-1 and 8-to-CCZ protocol \cite{bravyi2005universal, Litinski_2019}, $m = 5$ ($m = 4$ for 8-to-CCZ)  fits comfortably into each six-qubit logical block of the 12-qubit Gross and two-Gross codes. This enables a natural \emph{dual-track} mode: run two copies of the protocol simultaneously on the two ZX-dual blocks, effectively doubling factory throughput without adding code patches.

As discussed in Section~\ref{sec:simultaneous-measure}, qubits $L_0$ and $L_6$ form a dual pair under ZX-duality, and the LPU is attached to both. The LPU also decomposes into distinct $X$ and $Z$ modules. When both modules operate in the same basis, the architecture supports simultaneous $X$ or $Z$ measurements on $L_0$ and $L_6$. Because the automorphism group acts identically on the two six-qubit blocks, this parallelism extends to more general Pauli measurements, as long as the logical Paulis on the two blocks coincide and are purely $X$ or $Z$ type.

\begin{figure}[t]
    \centering
    \includegraphics[width=0.4\textwidth]{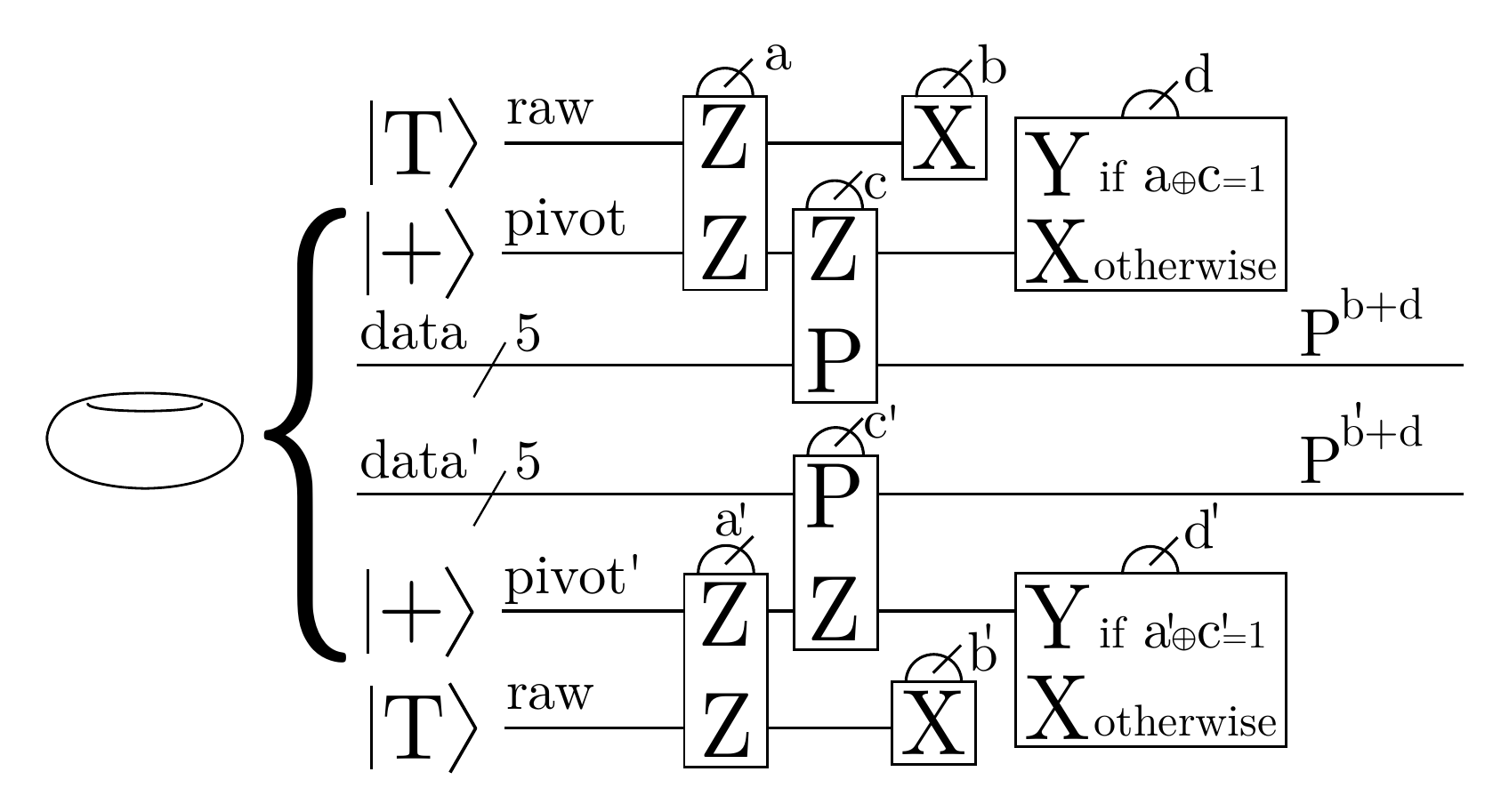}
    \caption{
    Simultaneous realization of two pivot-based injections on blocks $L_0$ to $L_5$ and $L_6$ to $L_{11}$. Logical Paulis on the two blocks are chosen to be identical and of pure $X$ or $Z$ type, which is essential for parallelization in \emph{dual-track} distillation. Only pivot $Y$ measurements, which require both LPU modules, must be serialized. Similar patterns apply to the other injection schemes in Figure~\ref{fig:magic_inject}.}
    \label{fig:simultaneous_inject}
\end{figure}

These properties align well with direct-injection schemes (Figure~\ref{fig:magic_inject}), see Fig.~\ref{fig:simultaneous_inject}. For most rotation steps, we can schedule paired measurements on the two blocks with identical logical labels so that one sequence of automorphisms followed by a simultaneous LPU measurement implements both rotations. This yields a near factor-of-two throughput improvement for the same LPU footprint.

The main exception occurs at steps that require $Y$-basis measurements on the pivot. A $Y$ measurement occupies both the $X$ and $Z$ modules, so the two protocol copies must serialize at those points. In pivot-based injection, the need for a pivot $Y$ measurement is tied to whether a correction is required, which happens with probability $3/4$. In these cases, multi-track execution does not reach a strict factor-of-two speedup, but still provides a significant throughput gain, especially when the protocol is dominated by $X$ and $Z$ rotations and when direct injection reduces pivot usage.

\section{Protocol-level Compression of Distillation Footprint: Recycling Logical Qubits}
\label{sec:compression}
\begin{figure*}[t]
    \centering
    \includegraphics[trim=0 0.5in 0 0, width=\textwidth]{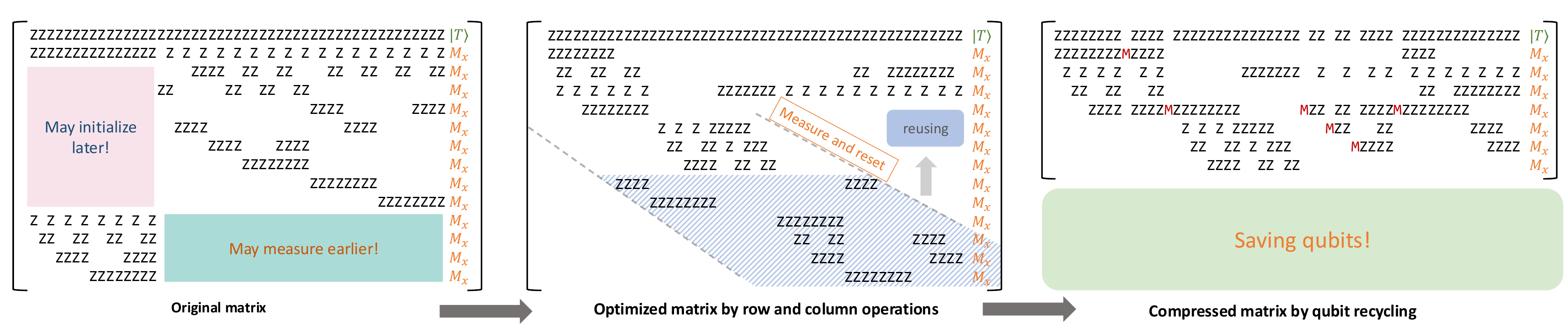}
    \caption{
    (a) Example of a 49-to-1 magic state distillation triorthogonal matrix showing two all-zero subblocks: the left (pink) region corresponds to rows that can be initialized later, and the right (green) region to even rows that can be measured earlier. (b) After row and column operations that preserve triorthogonality, logical qubits freed by early-measured rows are recycled to support later-initialized rows, yielding a compressed matrix with fewer simultaneously active rows. (c) The optimized matrix achieves substantial logical-qubit savings while preserving protocol correctness.}
    \label{fig:compress}
\end{figure*}

The limited number of logical qubits in BB codes makes it challenging to host large distillation protocols within a single patch. Instead, we reduce the protocol's logical-qubit footprint by exploiting structure in its triorthogonal matrix. This optimization does not rely on implementation-specific details of the distillation circuits, and therefore applies to any distillation protocol generated from triorthogonal matrices, including protocols for states such as $\ket{T}$, $\ket{CS}$, and $\ket{CCZ}$.

We introduce a \emph{protocol compression} technique that lowers the peak number of simultaneously active logical qubits by recycling qubits whose rows become idle. In many triorthogonal constructions, such as the 49-to-1 code of Bravyi and Haah \cite{Bravyi_2012}, the matrix contains large all-zero subblocks that indicate opportunities to delay initialization or advance measurement. 

Let $G \in \{0,1\}^{m \times n}$ be a binary triorthogonal matrix. The first $k$ rows have odd Hamming weight and encode the $k$ output logical qubits; the remaining $m - k$ rows have even weight. For each row $i$, let $f_i$ denote the index of its first $1$ (or $+\infty$ if the row is all zeros) and $\ell_i$ the index of its last $1$ (or $-\infty$ if all zeros).

Even rows support stabilizer checks and can both start later and end earlier: if the leftmost nonzero entry is at column $f_i$, the row need not be initialized before column $f_i$, and if the rightmost nonzero entry is at $\ell_i$, the row can be measured and freed after column $\ell_i$. Odd rows, in contrast, encode outputs and cannot be freed once initialized; even if an odd row has trailing zeros, we treat it as active on all columns $j \ge f_i$.

We say row $i$ is \emph{working} on column $j$ if $j \ge f_i$ and either  
(i) $j \le \ell_i$ and the row is even, or  
(ii) the row is odd.  
Let $W(j)$ denote the set of working rows at column $j$. The required number of simultaneous logical qubits is
\[
    \mathcal{C}(G) = \max_{j \in [n]} |W(j)|.
\]
Our goal is to transform $G$ into an equivalent triorthogonal matrix $G'$ with the same distillation properties but a smaller peak footprint $\mathcal{C}(G')$.

We allow three classes of transformations that preserve triorthogonality and the encoded protocol:
\begin{itemize}
    \item Column permutations, which reorder the commuting $\pi/8$ rotations.
    \item Row permutations within blocks, which reorder odd rows among themselves and even rows among themselves.
    \item Row additions over $\mathbb{F}_2$, which add one row to another while maintaining triorthogonality and logical content.
\end{itemize}

By applying these operations, we reshape $G$ so that many even rows share a right-aligned all-zero submatrix (early measurement) and many rows share a left-aligned all-zero submatrix (delayed initialization). In terms of the intervals $[f_i,\ell_i]$, these transformations shorten active windows and reduce the maximum overlap $\max_j |W(j)|$.

Figure~\ref{fig:compress} illustrates this process for a 49-to-1 protocol. The original matrix has prominent left- and right-aligned all-zero regions. After suitable row additions and permutations, qubits freed by early-measured rows are recycled to support later-initialized rows. The resulting compressed matrix has a significantly reduced peak logical-qubit footprint while preserving triorthogonality and output error suppression.

Finding the globally optimal compression is computationally hard. Even in the simplified case where $k = 0$ and each even row has Hamming weight two, minimizing $\mathcal{C}(G)$ reduces to the NP-hard cutwidth problem. For realistic protocols with $n$ in the tens or higher, we therefore rely on heuristics. In our experiments, simple greedy schemes that cluster row starts and ends, combined with targeted row additions, already yield substantial qubit savings and allow otherwise infeasible protocols to fit within a single BB patch.

\section{Evaluation Methodology}
\label{sec:eval}

\subsection{Baselines}
\label{subsec:baseline}
We compare our proposed distillation factories (\textsc{gross} and \textsc{two-gross}) against two state-of-the-art magic-state factory baselines: a surface-code distillation baseline follows the lattice-surgery factories of Litinski~\cite{Litinski_2019}, and a cultivation baseline from Gidney’s grafted surface code magic state cultivation~\cite{gidney2024magicstatecultivationgrowing}.
Factories are evaluated under two different physical error rates \(p_{\mathrm{phys}}\), with various input magic state error rates \(p_{\mathrm{in}}\) and output magic state error rates \(p_{\mathrm{out}}\).
Each factory is characterized by its physical-qubit footprint, the number of logical timesteps \(\tau_i\) per batch, and the resulting space-time volume (qubits \(\times\) timesteps). These are the quantities reported and compared in our results.

Distillation baselines are labeled as \((\mathrm{Protocol})_{\mathrm{Code}}\), and, when the protocol is implemented on the surface code, we further annotate a triple \((d_X, d_Z, d_m)\) that specifies the code distances used for the data blocks in~\cite{Litinski_2019}. 
Cultivation baselines are written as \((\mathrm{Cultivation})_{\mathrm{SC}\rightarrow d}\), where \(d\) is the distance of Gidney’s grafted surface-code patch.

\subsection{Factory Usage Modes}\label{sec:our-factory}
Magic state distillation protocols require a source of raw magic states as input. Our protocols then operate on magic states loaded into logical qubits of a bivariate bicycle code. We consider two settings in which our methods can be applied:

\textbf{Two-round distillation.} 
The bicycle architecture \cite{yoder2025tour} details a promising approach for obtaining magic states by connecting magic state cultivation protocols \cite{gidney2024magicstatecultivationgrowing} to a BB memory via a surgery ancilla system called \textit{adapter} \cite{swaroop2025universaladaptersquantumldpc}.
Magic state cultivation is a compressed hardware-native protocol that can achieve error rates as low as $10^{-9}$ to $10^{-11}$ depending on the details of the construction. 
Novel designs of magic state cultivation are still emerging \cite{sahay2025foldtransversalsurfacecodecultivation} and may need to be tailored to hardware limitations, but the adapter construction is flexible enough such that any such proposal could be integrated into a bicycle architecture. 
In this setting, cultivation would act as a first-round protocol, with our distillation protocols implemented in BB memory acting as a second-round protocol achieving suppressed error rates (e.g., $35\cdot(10^{-9})^{3} = 3.5 \cdot 10^{-26}$ with a $10^{-9}$ cultivated state fed into a $[[15,1,3]]$ protocol). 
We also consider more conventional factory designs, where the first and second round protocols are both distillation.
In our later results, such configurations are written as a combination of the two round protocols to make the first- and second-round costs explicit.

\textbf{One-round distillation.} 
While magic state cultivation is an optimized, high-performing protocol, we also consider other first-round protocols for adaptability to different architectural settings. 
An emerging line of work \cite{Li_2015, xu2025batchedhighratelogicaloperations, yoshioka2025transversalgatesprobabilisticimplementation} is considering lower-overhead methods that instead inject low-quality magic states (physical error rates $10^{-2}$ to $10^{-3}$) into the memory directly, in which case our methods would act as a first round of error suppression. 
These could then be used for applications of early fault-tolerant scale (e.g., $35\cdot(10^{-3})^{3} = 3.5 \cdot 10^{-8}$ with a $10^{-3}$ raw state in a $[[15,1,3]]$ protocol).

\subsection{Noise Model and Error Analysis}\label{sec:factory-error-analysis}
A crucial part of magic-state factory design is understanding how each error source affects the final output error rate and the discard (post-selection) rate. Here, we present a detailed noise model and error analysis for our implementation of the distillation circuit in the BB architecture. 
We explicitly model logical errors, including imperfect logical operation and imperfect measurement outcomes. 
For both the gross code and the two-gross code, we characterize the combined impact on the delivered state’s fidelity and acceptance probability for the following noise resources:

\begin{enumerate}
  \item \textbf{Magic state input error} $p_{\textrm{in}}$.
  This error arises from imperfect preparation of input $T$ magic states.
  We model it as a depolarizing channel applied to the ideal magic state $\ket{m}$.
  The resulting error should be suppressed polynomially by the distillation protocol.

  \item \textbf{Automorphism gate error} $p_{\textrm{auto}}$.
  Automorphism errors are introduced by physical \texttt{CNOT} gates (each with physical error rate $p_{\textrm{phys}}$) applied to data and ancilla qubits.
  We model these gate errors as a uniform depolarizing channel acting on each logical qubit.

 \item \textbf{Inter-module measurement error} $p_{\textrm{inter}}$.
  This error is introduced by the adapter that connects the distillation BB code patch with the magic state input. In this paper, we assume such error is concentrated on the two qubits measured, modeled as two-qubit depolarizing logical errors. In the bypassing pivot injection scheme in Figure ~\ref{fig:magic_inject}(a)(c), the magic state is injected directly into logical qubits of the distillation circuit, so the logical error channel appears in the injection gadget as a multi-qubit depolarizing channel with inter-module logical error rate. While in the pivot injection scheme in Figure ~\ref{fig:magic_inject}(b), Pauli errors introduced by the inter-module measurement are modeled as a two-qubit depolarizing channel acting on the source and pivot qubits, and are then propagated to the injected $T$ magic state at the pivot, though at the cost of additional measurement operations. Below we detail how these errors propagate from the magic state resource with a concrete example.

  \item \textbf{In-module measurement error} $p_{\textrm{intra}}$.
   This error is introduced by the LPU, which is used for logical operations including magic state injection, correction, and final post selection. The total error rate contains two components: measurement error and logical qubit error. The former correspond to flipping of measurement results, and the latter correspond to depolarizing logical errors. We define $\lambda$ as the ratio between measurement error and the total in-module error rate, $\lambda \;=\; \frac{p_{\mathrm{meas}}}{p_{\textrm{intra}}}\, .$ 
\end{enumerate}

\begin{figure}[t]
         \centering
         \includegraphics[width=0.48\textwidth]{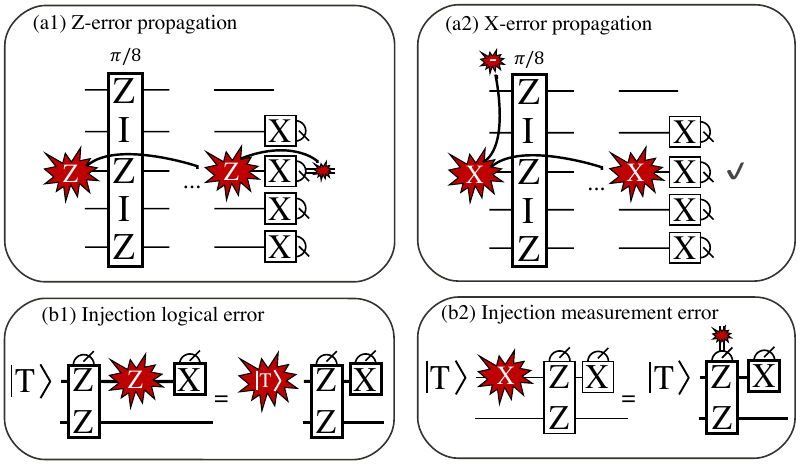}
         \caption{Illustration of how different errors are handled in our simulations. (a1) A Pauli-$Z$ logical error leaves rotations unaffected but may flip final parity checks. (a2) A Pauli-$X$ error flips the sign of rotations, but leaves final measurements unaffected. (b1) A faulty in-module/inter-module measurement introduces logical depolarizing errors to qubits, and (b2) a fault measurement outcome can be interpreted as a faultier input magic state.}
         \label{fig:errormodel}
\end{figure}

Not all logical errors harm the output magic state~\cite{Litinski_2019}. Figure~\ref{fig:errormodel} gives examples of how faults from logical operations propagate through the distillation circuit and contribute to the output error. In Fig.~\ref{fig:errormodel}(a1), a Pauli \(Z\) logical error on qubits \(2\) to \(5\) commutes through all rotations and is detected by the final detector. In contrast, a Pauli \(X\) error in Fig.~\ref{fig:errormodel}(a2) flips every measurement outcome it meets and is invisible to the final detector. Even in this case, it does not necessarily cause an output error, since at least three flipped rotations are required to induce a logical fault; the combination of the \(X\)-error location and the gate schedule is therefore crucial.

In Fig.~\ref{fig:errormodel}, logical measurement errors are split into the two models introduced above. 
Panel (b1) models a two-qubit depolarizing channel acting on the measured qubits for inter-module measurements and all qubit depolarizing channel for in-module measurements.
Specifically, the error on the magic state qubit is less harmful, as any \(Z\) component is equivalent to a \(Z\) preparation error that introduces an extra \(\pi/2\) rotation, while an \(X\) component can be absorbed into the final \(X\)-basis measurement. 
Panel (b2) illustrates a pure measurement-outcome flip, which is equivalent to inserting (or omitting) a \(\pi/4\) rotation (or an input \(X\) error on the magic state) and is detectable by the distillation protocol. 
Finally, errors in the final \(X\)-basis measurements at the end are generally less harmful as well: a false positive only increases the discard rate, while a false negative must combine with a preexisting logical error and is therefore a second-order contributor to the final output error rate.

Hence, we report the output error rate by two methods: (i) a union bound calculation, which counts any failure from any logical operation toward the output infidelity; and (ii) a density matrix simulation that models each logical qubit as a single qubit with the prescribed logical level noise rates. In the simulation, parameter $\lambda$ sets the mix of logical measurement errors (measurement flips versus depolarizing memory faults). Both methods employ error rates from previous simulation results in Table~\ref{tab:errorrate} of~\cite{yoder2025tour}. Notably, due to the highly non-Clifford nature of the distillation circuits considered here, end-to-end physical-level simulation is computationally impractical in the regime of interest. Exact simulation of generic noisy non-Clifford circuits typically requires dense density-matrix methods rather than efficient stabilizer-based techniques, and is therefore in practice restricted to only tens of qubits, often on the order of $\sim20$ qubits. Moreover, a full physical-level treatment would need to incorporate repeated syndrome extraction, decoding, and memory noise over time in order to determine the effective logical error rates of the BB operations. Our two-level methodology instead uses prior physical-level studies from \cite{yoder2025tour} determine these hardware-informed logical error rates, and then evaluates the performance of the full distillation circuit at the logical level under the resulting noise model.

\subsection{Experimental Setup and Classical Compilation Overhead}
All simulations were conducted on a Macbook Pro with a 10 core CPU and 32 GB of RAM. We also report the classical running time of each optimization in Table~\ref{tab:calssical}, using the example of the largest protocol 49-to-1 and largest code two-gross.

\begin{table}[t]
\setlength{\tabcolsep}{5pt}
\scriptsize
\centering
\begin{tabular}{@{}c||c|c|c|c@{}}
\hline\hline
 \multirow{2}{*}{Logical Operation} & \multirow{2}{*}{ Code Type} &  \multirow{2}{*}{\makecell[c]{Timesteps \\ $\tau_i$}}  & \multicolumn{2}{c}{Logical Error Rate $P$} \\
 \cline{4-5}
  &  &    & $p_\textrm{phys}=10^{-3}$& $p_\textrm{phys}=10^{-4}$  \\
\hline
\hline
 \multirow{2}{*}{\makecell[c]{Shift automorphism \\ $p_\textrm{auto}$}} & Gross    & $14$ & $10^{-6.4}$  & $10^{-12.2}$ \\
                     & Two-gross & $14$ & $10^{-14.5}$ & $10^{-37}$     \\
\hline
 \multirow{2}{*}{\makecell[c]{In-module meas. \\ $p_\textrm{intra}$}} & Gross     & $120$ & $10^{-5.0}$ & $10^{-9.0}$ \\
                & Two-gross & $216$ & $10^{-11}$      & $10^{-20}$      \\
\hline
 \multirow{2}{*}{\makecell[c]{Inter-module meas. \\ $p_\textrm{inter}$}} & Gross   & $120$ & $10^{-2.7}$ & $10^{-7.3}$ \\
                     & Two-gross & $216$ & $10^{-9}$       & $10^{-18}$      \\
\hline\hline
\end{tabular}
\caption{Timestep and logical error rate of BB code’s logical operations from~\cite{yoder2025tour}. The tabulated values are identical to those employed in our simulations.}
\label{tab:errorrate}
\end{table}

\begin{table}[t]
\setlength{\tabcolsep}{5pt}
\scriptsize
\centering
\begin{tabular}{l r}
\hline
Optimization pass & Runtime \\
\hline
Logical qubit mapping \& masking & $\sim 10\,\mathrm{min}$ \\
Automorphism TSP & $< 3\,\mathrm{s}$ \\
Protocol compressor & $< 5\,\mathrm{s}$ \\
\hline
\end{tabular}
\caption{Classical Compilation Overhead of the Optimization Methods in Sec.~\ref{sec:optimization} and \ref{sec:compression} for two-gross code and 49-to-1 protocol.}
\label{tab:calssical}
\end{table}

\section{Results}
\label{sec:results}
\subsection{Resource Estimation and Comparison}

\begin{table*}[t]
\setlength{\tabcolsep}{4pt}
\scriptsize
    \centering
    \begin{tabular}{ c||c|c|c|c|c|c|c|c|c}
    \hline\hline
    \multirow{2}{*}{One-Round Factories} &
    \multirow{2}{*}{$p_{\textrm{phys}}$} &
    \multicolumn{2}{c|}{\multirow{2}{*}{$p_{\textrm{in}}$}} &
    \multirow{2}{*}{Physical Qubits} &
    \multirow{2}{*}{\makecell[c]{Timesteps \\ $\tau_i$}} &
    \multirow{2}{*}{\makecell[c]{Space-time \\ Volume}} &
    \multirow{2}{*}{\makecell[c]{Union Bound}} &
    \multicolumn{2}{c}{Simulated} \\
    \cline{9-10}
     && \multicolumn{2}{c|}{}& & & & &
     \makecell[l]{$p_\textrm{out}$ ($\lambda = 0.9$)} &
     \makecell[l]{$p_\textrm{out}$ ($\lambda = 0.5$)} \\
     \hline 
     \hline
     $\textrm{(15-to-1)}_{\textrm{SC(17,7,7)}}$\cite{Litinski_2019} 
       & $10^{-3}$ & \multicolumn{2}{c|}{$10^{-3}$} 
       & $4620$  & $256$ & ~$1.2\times 10^6$ & N/A 
       & \multicolumn{2}{c}{$4.5 \times10^{-8}$}  \\
     \hline
     $\textrm{(Cultivation)}_{\textrm{SC} \rightarrow\textrm{d=3}}$ \cite{gidney2024magicstatecultivationgrowing,yoder2025tour}
       & $10^{-3}$ & \multicolumn{2}{c|}{$10^{-3}$} 
       & $454$  & $351$ & ~$1.6\times 10^5$ &  N/A 
       & \multicolumn{2}{c}{$3 \times10^{-6}$ } \\
     \hline
     $\textrm{(Cultivation)}_{\textrm{SC} \rightarrow\textrm{d=5}}$ \cite{gidney2024magicstatecultivationgrowing,yoder2025tour} 
       & $10^{-3}$ & \multicolumn{2}{c|}{$10^{-3}$} 
       & $463$  & $2167$ & ~$1.0\times 10^6$ &  N/A 
       & \multicolumn{2}{c}{$2\times10^{-9}$}  \\
     \hline
      $\textrm{(8-to-CCZ)}^{\otimes 2}_{\textrm{Gross}}$
       & $10^{-3}$ & \multicolumn{2}{c|}{$10^{-3}$} 
       & $378$  & $1570$ & $5.9\times 10^5$ & $3.2 \times 10^{-4}$ 
       & $8.8 \times 10^{-5}$  & $9.8 \times 10^{-5}$ \\
     \hline
     $\textrm{(15-to-1)}_{\textrm{Gross}}$
       & $10^{-3}$ & \multicolumn{2}{c|}{$10^{-3}$} 
       & $378$  & $6122$ & $2.3\times 10^6$ & $5.0\times 10^{-4}$ 
       & $ 1.3 \times10^{-6}$ & $ 4.6 \times10^{-6}$   \\
     \hline
     $\textrm{(20-to-4)}_{\textrm{Gross}}$
       & $10^{-3}$ & \multicolumn{2}{c|}{$10^{-3}$} 
       & $378$  & $3088$ & $1.2\times 10^6$ & $9.8 \times 10^{-4} $ 
       & $ 4.3 \times10^{-5}$ & $ 5.2 \times10^{-5} $   \\
     \hline
     $\textrm{(15-to-1)}_{\textrm{Two-gross}}$
       & $10^{-3}$ & \multicolumn{2}{c|}{$10^{-3}$} 
       & $734$  & $11249$ & $8.3\times 10^6$ & $1.1\times10^{-8}$ 
       & $1.0\times10^{-8}$  & $1.0\times10^{-8}$   \\
     \hline
     $\textrm{(49-to-1)}_{\textrm{Two-gross}}$
       & $10^{-3}$ & \multicolumn{2}{c|}{$10^{-3}$} 
       & $734$  & $70748$ & $5.1\times 10^7$ & $ 3.1 \times 10^{-9}$ 
       & $2.0 \times 10^{-11} $  & $9.7 \times 10^{-11} $   \\
     \hline
     $\textrm{(15-to-1)}_{\textrm{SC(11,5,5)}}$\cite{Litinski_2019} 
       & $10^{-4}$ & \multicolumn{2}{c|}{$10^{-4}$} 
       & $2070$  & $180$ & ~$3.7\times 10^5$ &  N/A 
       & \multicolumn{2}{c}{$1.9 \times10^{-11}$}  \\
     \hline
     $\textrm{(15-to-1)}_{\textrm{Gross}}$
       & $10^{-4}$ & \multicolumn{2}{c|}{$10^{-4}$} 
       & $378$  & $5999$ & ~$2.3\times 10^6$ & $5.1 \times10^{-8}$  
       & $9.4 \times10^{-11}$ & $4.2 \times10^{-10}$ \\
     \hline
     $\textrm{(15-to-1)}_{\textrm{Two-gross}}$
       & $10^{-4}$ & \multicolumn{2}{c|}{$10^{-4}$} 
       & $734$  & $11090$ & $8.1\times 10^6$ & $1.0 \times10^{-11}$ 
       & $1.0 \times10^{-11}$ & $1.0 \times10^{-11}$ \\
     \hline
     $\textrm{(49-to-1)}_{\textrm{Two-gross}}$
       & $10^{-4}$ & \multicolumn{2}{c|}{$10^{-4}$} 
       & $734$  & $68595$ & $5.0\times 10^7$ & $ 1.2 \times 10^{-17}$ 
       & $\leq 10^{-17}$  & $\leq 10^{-17}$   \\
     \hline\hline
    \multirow{2}{*}{Two-Round Factories} &
    \multirow{2}{*}{$p_{\textrm{phys}}$} &
    \multicolumn{2}{c|}{$p_{\textrm{in}}$} &
    \multirow{2}{*}{\makecell[c]{Physical Qubits \\ \nth{1}+\nth{2} Round}} &
    \multirow{2}{*}{\makecell[c]{Timesteps \\ $\tau_i$}} &
    \multirow{2}{*}{\makecell[c]{\nth{2} Round\\ Volume}} &
    \multirow{2}{*}{\makecell[c]{Union Bound}} &
    \multicolumn{2}{c}{Simulated} \\
    \cline{9-10}
    \cline{3-4}
     & &\nth{1} &\nth{2} & & & & &
     \makecell[l]{$p_\textrm{out}$ ($\lambda = 0.9$)} &
     \makecell[l]{$p_\textrm{out}$ ($\lambda = 0.5$)} \\
     \hline 
     \hline
     \makecell[l]{$\textrm{(15-to-1)}_{\textrm{SC(11,5,5)}} + \textrm{(15-to-1)}_{\textrm{SC(25,11,11)}}$ \cite{Litinski_2019} } 
       &  $10^{-3}$ & $10^{-3}$ & N/A
       & \makecell[c]{$12420 + 18280$} & $495$ & ~$9.1\times 10^6$ &  N/A 
       & \multicolumn{2}{c}{ $2.7 \times10^{-12}$}\\
     \hline
     \makecell[l]{$\textrm{(Cultivation)}_{\textrm{SC}}$\cite{gidney2024magicstatecultivationgrowing,yoder2025tour}$+ \textrm{(15-to-1)}_{\textrm{Two-gross}}$} 
       &  $10^{-3}$ & $10^{-3}$ & $10^{-6}$
       & $454+734$  & $11080$ & $8.1\times 10^6$ & $ 4.0\times10^{-10}$ 
       & $8.2 \times10^{-13}$ & $4.1 \times10^{-12}$\\ 
     \hline
     \makecell[l]{ $\textrm{(15-to-1)}_{\textrm{SC(7,3,3)}} + \textrm{(8-to-CCZ)}_{\textrm{SC(15,7,9)}}$\cite{Litinski_2019} } 
       &  $10^{-4}$ & $10^{-4}$ & $10^{-7.36}$  
       & $3240+9160$ & $217$ & ~$2.0\times 10^6$ &  N/A 
       & \multicolumn{2}{c}{ $7.2 \times10^{-14}$}\\
     \hline
     \makecell[l]{$\textrm{(15-to-1)}_{\textrm{SC(7,3,3)}} + \textrm{(20-to-4)}_{\textrm{SC(13,5,7)}}$ \cite{Litinski_2019} }
       &  $10^{-4}$ & $10^{-4}$ & $10^{-7.36}$ 
       & $3240 + 8340$ & $420$ & ~$3.5\times 10^6$ &  N/A 
       & \multicolumn{2}{c}{ $1.4 \times10^{-12}$}\\
     \hline
     \makecell[l]{ $\textrm{(15-to-1)}_{\textrm{SC(7,3,3)}}$ \cite{Litinski_2019}   $+ \textrm{(8-to-CCZ)}^{\otimes 2}_{\textrm{Two-gross}}$}
       &  $10^{-4}$ & $10^{-4}$ & $10^{-7.36}$
       & $1620+734$  & $2893$ & $2.1\times 10^6$ & $ 2.4\times10^{-14}$ 
       & $2.4 \times10^{-14}$ & $2.4 \times10^{-14}$\\ 
     \hline
     \makecell[l]{$\textrm{(15-to-1)}_{\textrm{SC(7,3,3)}}$ \cite{Litinski_2019}    $+ \textrm{(20-to-4)}_{\textrm{Two-gross}}$  } 
       &  $10^{-4}$ & $10^{-4}$ & $10^{-7.36}$ 
       & $810+734$  & $5235$ & $3.8\times 10^6$ & $ 1.9\times10^{-14}$ 
       & $1.1 \times10^{-14}$ & $1.1 \times10^{-14}$\\ 
     \hline
     \makecell[l]{$\textrm{(Cultivation)}_{\textrm{SC}}$\cite{gidney2024magicstatecultivationgrowing,yoder2025tour} $+ \textrm{(15-to-1)}_{\textrm{Two-gross}}$} 
       &  $10^{-4}$ & $10^{-4}$ & $10^{-6}$
       & $454+734$  & $11080$ & $8.1\times 10^6$ & $1.0 \times10^{-17}$ 
       & $\leq 10^{-17}$ & $\leq 10^{-17}$ \\ 
     \hline
    \end{tabular}
    \caption{Resource comparison between different magic state factories across different distillation protocols using pivot-injection. For each factory we list the physical error rate $p_{\mathrm{phys}}$, input magic-state error $p_{\mathrm{in}}$, physical qubit footprint, timestep depth $\tau_i$ (discard rate counted), the resulting 
    space-time volume (qubits~$\times$~timesteps), and the output error $p_{\mathrm{out}}$ obtained from the analytic model and from density-matrix simulation; see Sec.~\ref{subsec:baseline} for definitions and the baseline configurations. Two-round factories are written as ``first-round~+ second-round'' and the reported resources correspond to the second round. Entries of the form $(\cdot)^{\otimes 2}$ denote dual-track factories, as described in Sec.~\ref{subsec:multi-track}. All benchmarks in this table adopt pivot injection in Figure~\ref{fig:magic_inject}(b) for better inter-module error suppression. Across the listed baselines, gross and two-gross factories often reduce physical-qubit footprint while trading off depth and, in some settings, output error.}
    \label{tab:resource}
\end{table*}

Table~\ref{tab:resource} summarizes the resource requirements of our magic-state factories across multiple distillation protocols, where output error rates are obtained from our logical-level simulations in Section~\ref{sec:eval} with logical error rates from Table~\ref{tab:errorrate}.

Unless explicitly stated otherwise, all numbers in Table~\ref{tab:resource} (Table II) and in the corresponding comparison plots use pivot injection in Fig.~\ref{fig:magic_inject}(b). The impact of injection choice is isolated in Table~\ref{tab:injectionmethods}, where the direct-injection family in Fig.~\ref{fig:magic_inject}(a),(c) is compared against pivot injection under the same factory settings.

At a physical error rate of \(p_{\text{phys}} = 10^{-3}\), gross-code distillation used as a one-round factory provides substantial qubit savings compared to surface-code baselines, at the cost of larger \(\tau_i\). For example, \(\textrm{(15-to-1)}_{\textrm{Gross}}\) uses 378 qubits versus 4620 for \(\textrm{(15-to-1)}_{\textrm{SC(17,7,7)}}\), while requiring larger depth/volume (\(6122,\ 2.3\times10^6\) vs \(256,\ 1.2\times10^6\)). Because native-measurement logical error in the gross code is relatively high, gross code is better matched to quadratic-suppression protocols (rows \(\textrm{(8-to-CCZ)}^{\otimes 2}_{\textrm{Gross}}\), \(\textrm{(20-to-4)}_{\textrm{Gross}}\)), whereas stronger suppression is obtained in higher-distance two-gross rows (\(\textrm{(15-to-1)}_{\textrm{Two-gross}}\), \(\textrm{(49-to-1)}_{\textrm{Two-gross}}\)).

While the cultivation scheme still has the lowest space-time overhead for target output error rates above about \(10^{-9}\), two-gross factories can reach significantly lower output error: \(\textrm{(49-to-1)}_{\textrm{Two-gross}}\) achieves \(10^{-11}\)-level output for 734 qubits at \(p_{\text{phys}}=10^{-3}\), and \(\le 10^{-17}\) at \(p_{\text{phys}}=10^{-4}\). Its second-round footprint is over \(20\times\) smaller than the second-round surface-code stage in \(\textrm{(15-to-1)}_{\textrm{SC(11,5,5)}}+\textrm{(15-to-1)}_{\textrm{SC(25,11,11)}}\) (734 vs 18280).

We also benchmark several two-round factories, analogous to the single-round case. At \(p_{\text{phys}}=10^{-3}\), \(\textrm{(Cultivation)}_{\textrm{SC}}+\textrm{(15-to-1)}_{\textrm{Two-gross}}\) reaches \(10^{-12}\)-level output error with comparable second-round volume to \(\textrm{(15-to-1)}_{\textrm{SC(11,5,5)}}+\textrm{(15-to-1)}_{\textrm{SC(25,11,11)}}\) (\(8.1\times10^6\) vs \(9.1\times10^6\)). At \(p_{\text{phys}}=10^{-4}\), the same hybrid reaches \(\le 10^{-17}\), compared with \(7.2\times10^{-14}\) for \(\textrm{(15-to-1)}_{\textrm{SC(7,3,3)}}+\textrm{(8-to-CCZ)}_{\textrm{SC(15,7,9)}}\), but with a larger second-round volume (\(8.1\times10^6\) vs \(2.0\times10^6\)). These row-to-row comparisons show that BB-based second rounds are particularly effective when very low output error is the primary target; because prior surface-code references provide only limited directly comparable baselines (without a fully matched all-surface cultivation+distillation hybrid), we report second-round volume for two-round rows.

Taken together, our results indicate that MSD on BB codes is a compelling building block for multi-round magic state factories.
We further note that the simulated error rates of the LPU measurements from~\cite{yoder2025tour} are likely to improve in the future, which means the performance of our proposed factories will improve as well.

\subsection{Expanding Native Measurements by Logical Mapping and Masking}
\begin{figure}[t]
         \centering
         \includegraphics[width=0.48\textwidth]{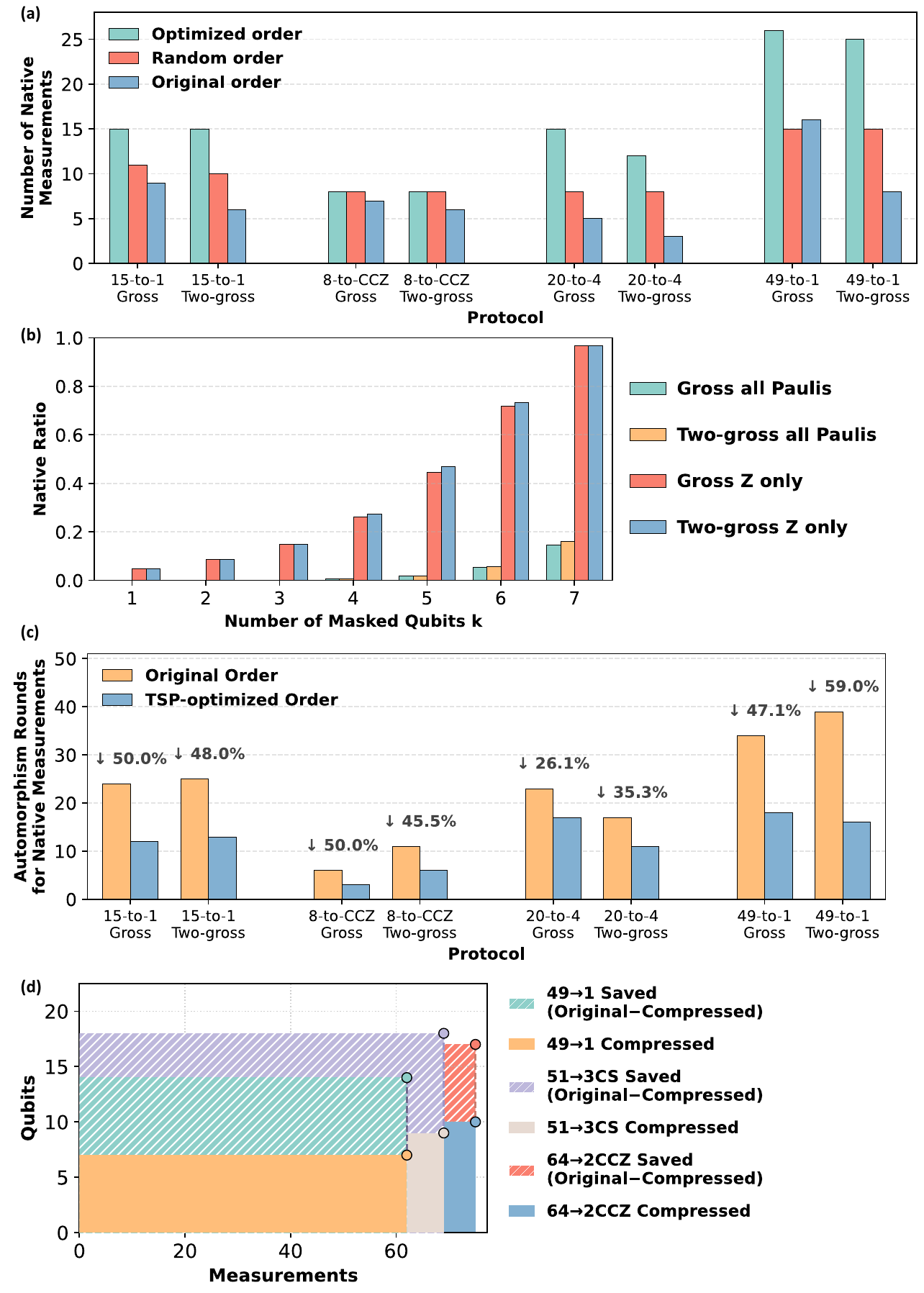}
         \caption{(a) Increase in native measurements from co-optimizing logical-qubit mapping and masking. ``Original order'' uses the circuit-derived mapping, while ``random order'' samples 10 random candidates and reports the best. (b) Native-measurement ratio versus masking in gross and two-gross codes, for both full-Pauli and $I/Z$-only operator sets. (c) Reduction in automorphism rounds from TSP-based scheduling, relative to the original gate order in \cite{Litinski_2019}. (d) Qubit-recycling compression reduces logical-qubit counts for large protocols without increasing total measurements.}
         \label{fig:natandcomp}
\end{figure}

As discussed in Sec.~\ref{subsec:logicalmapping} and~\ref{subsec:masking}, logical qubit mapping and masking increase native measurements in the gross and two-gross codes. Figure~\ref{fig:natandcomp}(a) compares mapping orders and shows that our search-based mapper outperforms brute-force and randomized baselines, making the 15-to-1 and 8-to-CCZ protocols fully native in both codes. Figure~\ref{fig:natandcomp}(b) shows that the native-measurement ratio grows with the number of masked qubits. We define this ratio as the number of native measurements divided by the number of candidate operators, evaluated for two operator sets: the full Pauli set \(\{I,X,Y,Z\}\) and an \(I/Z\)-only set. Masking substantially increases the share of \(I/Z\)-only native measurements required by the magic-state distillation circuits, and the same technique also benefits other quantum circuits executed on the BB architecture when blocks are not fully occupied. The resulting increase in native measurements reduces the synthesis cost of arbitrary Clifford operations and, in turn, lowers the overall circuit depth.

\subsection{Reducing Automorphism Rounds Using TSP optimization}
We schedule the commuting \(\pi/8\) rotations by solving a TSP-style ordering problem over Pauli measurements in Sec.~\ref{subsec:tsp}.
Figure~\ref{fig:tspmask}(c) illustrates the key effect: instead of following the protocol’s original column order, we cluster successive rotations that require similar automorphism actions, which reduces retargeting overhead between measurements. This optimization lowers both latency and accumulated logical-operation error because each eliminated automorphism round removes additional gates and measurement opportunities for faults. All timestep counts \(\tau_i\) reported in Tables~\ref{tab:resource} and~\ref{tab:injectionmethods} include this optimized scheduling.

\subsection{Benchmarking Distillation Protocol Compressor}
As discussed in Sec.~\ref{sec:compression}, our qubit-recycling distillation-protocol compressor reduces the number of simultaneously active logical qubits without increasing the total number of measurements. Figure~\ref{fig:natandcomp}(b) shows its impact on the logical-qubit footprint of large protocols. In particular, we compress the 49-to-1 protocol from 13 to 7 logical qubits, the 51-to-3CS protocol from 18 to 9 logical qubits, and the 64-to-2CCZ protocol from 17 to 10 logical qubits. These reductions are critical and allow both protocols to fit within a single gross or two-gross factory, whose maximum capacity is 11 logical qubits under our pivot-injection schemes.

\subsection{Identifying Dominant Error Sources}
\begin{figure}[t]
         \centering
         \includegraphics[width=0.48\textwidth]{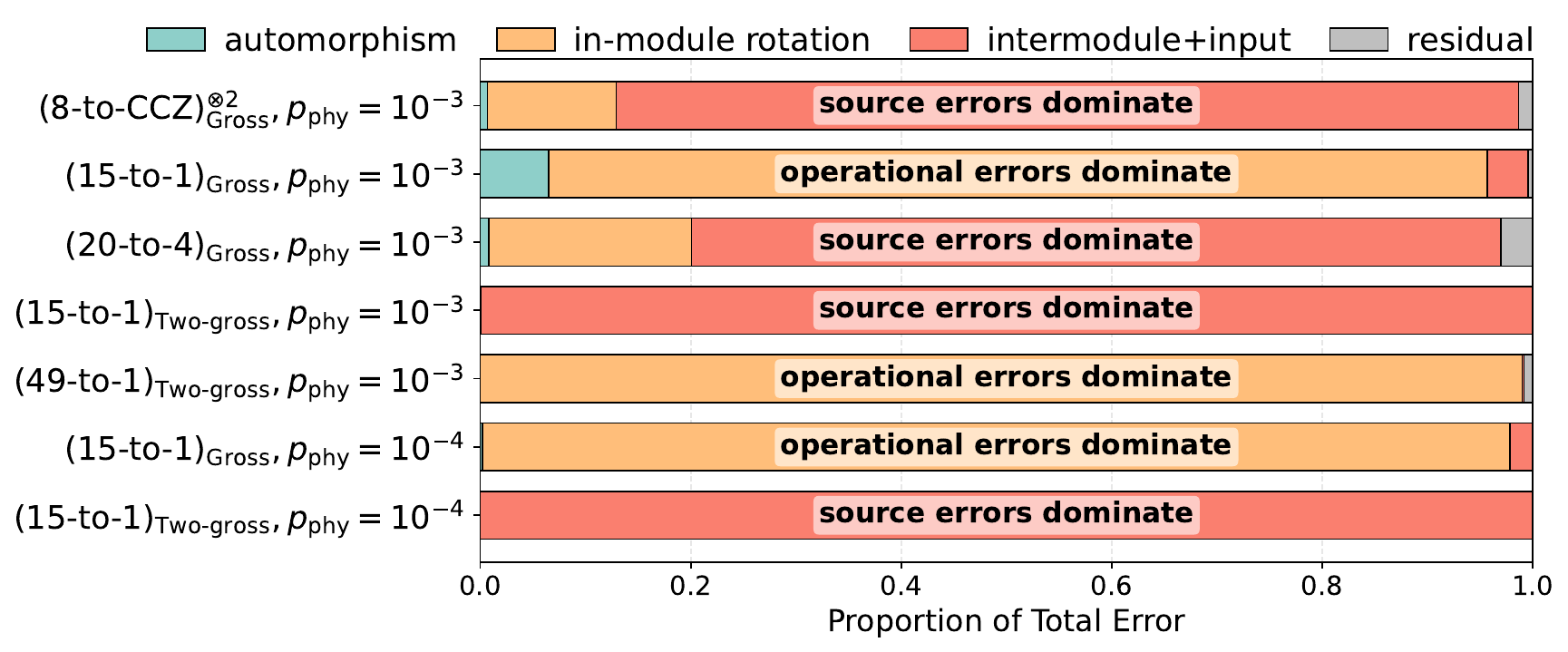}
         \caption{Dominant error source by distillation protocol. Most factories are limited either by \emph{operational errors} during distillation (e.g., in-module rotation errors) or \emph{source errors} (e.g., input magic state errors or inter-module injection errors).}
         \label{fig:errorcon}
\end{figure}

We decompose the output magic-state error of each simulated factory into contributions from three noise sources: (i) errors from automorphism gates, (ii) errors from in-module rotations, and (iii) errors in the injected magic states and inter-module measurements at the universal adapter, with Figure~\ref{fig:errorcon} summarizing the dominant source for each factory.
Most factories fall into one of two regimes: \emph{operation-limited}, where errors from in-module rotations dominate, and \emph{source-limited}, where imperfect input magic states and inter-module injection errors dominate. 
A practical design target is an operating point where source and operation errors contribute comparably to the final error; for example, at a physical error rate of \(p_{\text{phys}} = 10^{-3}\), the logical error of the gross code is too large to support the 15-to-1 protocol reliably, whereas the two-gross code is better matched to larger protocols such as 49-to-1. 
As logical error rates in the bicycle architecture improve with advances in LPU design and decoder optimization, this decomposition provides a compact indicator for tuning factory configurations to a given hardware error rate.

\subsection{Comparison Between Different Injection Schemes}
\label{subsec:differentinj}
\begin{table}[t]
\scriptsize
\centering

\setlength{\tabcolsep}{5pt}

\begin{tabular}{@{}c|c|c|c|c@{}}
\hline\hline
Factory & \makecell[c]{$p_{\textrm{phys}}=p_{\textrm{in}}$} & $\tau_i$& Space-time Volume &  $p_{\textrm{out}}^{\textrm{(sim)}}$ \\   
\hline
$\textrm{15-to-1}_{\textrm{gross}},{\textrm{direct}}$ 
  & $10^{-3}$ 
  & $2341$ 
  & $8.8 \times 10^{5}$ 
  & $8.2 \times 10^{-4}$  \\
\hline
$\textrm{15-to-1}_{\textrm{gross}},{\textrm{pivot}}$ 
  & $10^{-3}$ 
  & $6122$ 
  & $2.3 \times 10^{6}$ 
  & $4.6 \times 10^{-6}$  \\
\hline
$\textrm{15-to-1}_{\textrm{two-gross}},{\textrm{direct}}$ 
  & $10^{-3}$ 
  & $4145$ 
  & $3.0 \times 10^{6}$ 
  & $1.1 \times 10^{-8}$  \\
\hline
$\textrm{15-to-1}_{\textrm{two-gross}},{\textrm{pivot}}$ 
  & $10^{-3}$ 
  & $11249$ 
  & $8.3 \times 10^{6}$ 
  & $1.0 \times 10^{-8}$  \\
  \hline
$\textrm{8-to-CCZ}^{\otimes 2}_{\textrm{gross}},{\textrm{direct}}$ 
  & $10^{-3}$ 
  & $725$
  & $1.7 \times 10^{5}$ 
  & $2.5 \times 10^{-3}$  \\
\hline
$\textrm{8-to-CCZ}^{\otimes 2}_{\textrm{gross}},{\textrm{pivot}}$ 
  & $10^{-3}$ 
  & $1570$ 
  & $5.9 \times 10^{5}$ 
  & $9.8 \times 10^{-5}$  \\
\hline
$\textrm{49-to-1}_{\textrm{two-gross}},{\textrm{direct}}$ 
  & $10^{-3}$ 
  & $63651$ 
  & $4.7 \times 10^{7}$ 
  & $9.6 \times 10^{-9}$  \\
\hline
$\textrm{49-to-1}_{\textrm{two-gross}},{\textrm{pivot}}$ 
  & $10^{-3}$ 
  & $70748$ 
  & $5.2 \times 10^{7}$ 
  & $9.7 \times 10^{-11}$  \\
\hline
\hline
$\textrm{15-to-1}_{\textrm{gross}},{\textrm{direct}}$ 
  & $10^{-4}$ 
  & $2303$ 
  & $8.7 \times 10^{5}$ 
  & $2.1 \times 10^{-8}$  \\
\hline
$\textrm{15-to-1}_{\textrm{gross}},{\textrm{pivot}}$ 
  & $10^{-4}$ 
  & $5999$ 
  & $2.3 \times 10^{6}$ 
  & $4.2 \times 10^{-10}$  \\
\hline
\hline
\end{tabular}
\caption{Comparison between direct injection (Fig.~\ref{fig:magic_inject}(a),(c)) and pivot injection (Fig.~\ref{fig:magic_inject}(b)) across representative distillation protocols. Direct injection reduces distillation timesteps, while pivot injection suppresses low-fidelity inter-module measurement errors and typically achieves lower output error (detailed discussion in Sec.~\ref{subsec:differentinj}).}
\label{tab:injectionmethods}
\end{table}
Table~\ref{tab:injectionmethods} highlights a clear throughput-fidelity tradeoff between the two injection families. Direct injection is consistently faster because it avoids additional pivot measurements during T-state injection, but it inherits more logical memory errors from inter-module noise. Pivot injection adds logical measurements and therefore larger \(\tau_i\), yet confines inter-module noise to the source-pivot interface and yields better \(p_{\mathrm{out}}\) when inter-module measurements dominate.

This effect is strongest for factories composed entirely of native measurements. For example, the 15-to-1 gross-code factory at $p_{\mathrm{phys}}=10^{-3}$ improves from $8.2\times 10^{-4}$ with direct injection to $4.6\times 10^{-6}$ with pivot injection. For protocols requiring non-native rotations, such as 49-to-1, the depth advantage of direct injection is less consequential. In some two-gross cases, where inter-module measurements are cleaner, the performance gap narrows and the preferred choice becomes workload-dependent: direct injection is attractive for high throughput, while pivot injection remains preferable when targeting the lowest output error.

\subsection{Sensitivity Analysis in Logical Operation Error Rates}

Finally, we perform a one-factor-at-a-time sensitivity study around the baseline logical error rates in Table~\ref{tab:errorrate}. We exclude automorphism operations, since their error rates are primarily set by code design and are weakly sensitive to the specific LPU design. For each factory configuration, we sweep one logical operation error parameter at a time (in-module or inter-module measurement), fix the others, and record the resulting change in \(p_{\mathrm{out}}\).

\begin{figure}[t]
         \centering
         \includegraphics[width=0.48\textwidth]{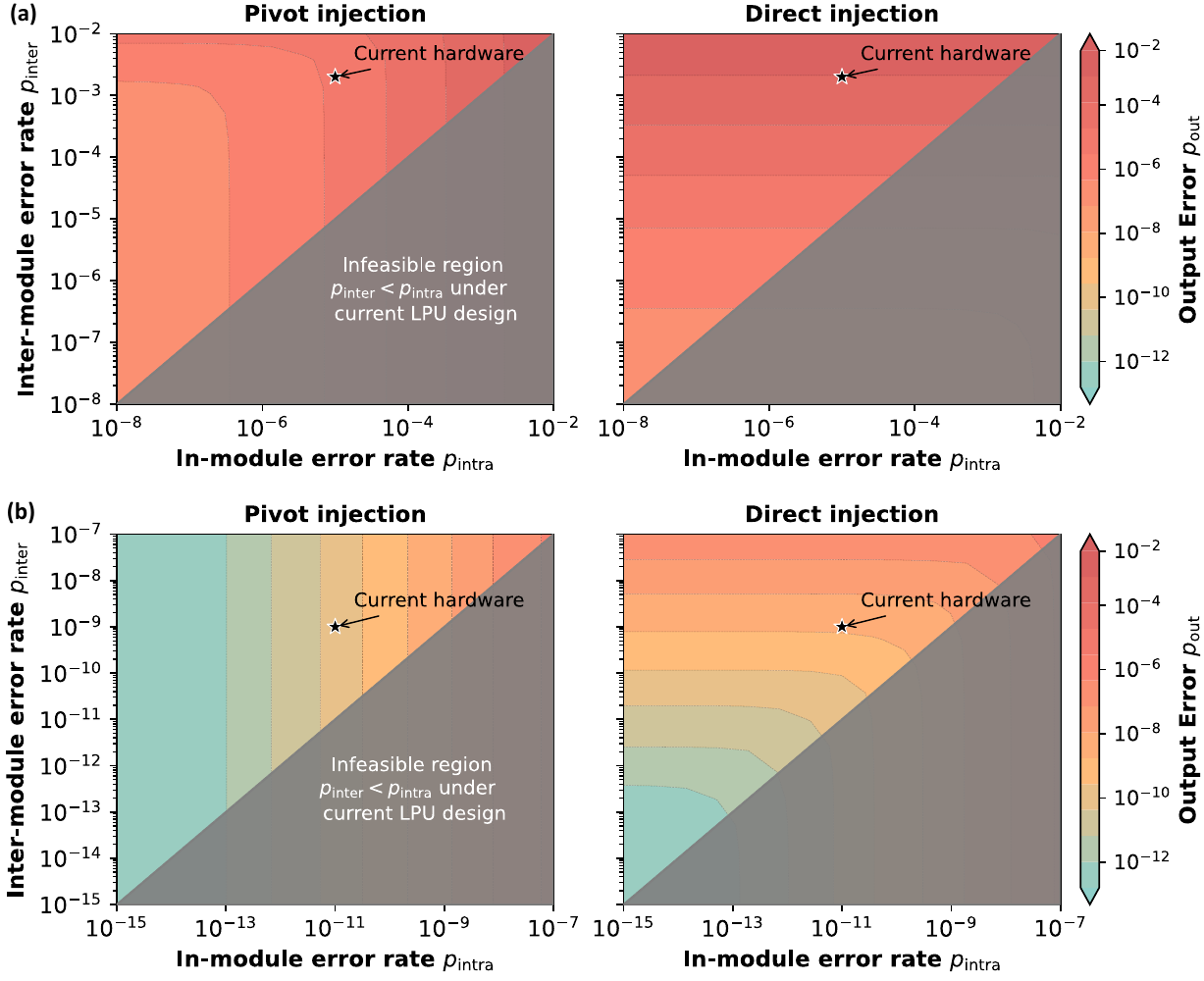}
         \caption{Sensitivity of output magic-state error to in-module and inter-module operation error rates. (a) 15-to-1 protocol for the gross code. (b) 49-to-1 protocol for the two-gross code. Since inter-module operations use in-module measurements as subroutines, we consider the regime $p_{\mathrm{inter}} \ge p_{\mathrm{intra}}$; the gray region indicates excluded parameters. The current-hardware point corresponds to logical error rates derived at physical error rate $p=10^{-3}$~\cite{yoder2025tour}.}
         \label{fig:sensitivity}
\end{figure}

Figure~\ref{fig:sensitivity} sweeps in-module and inter-module operation error rates and reports contours of constant output magic-state error. The knee of each contour marks the transition between in-module-dominated and inter-module-dominated regimes. Across the sweep, pivot injection is more robust to inter-module noise, shifting contours toward higher inter-module error rates at fixed output error. Overall, improving either error source helps, with the largest gains from better in-module measurement fidelity combined with pivot injection.

\section{Discussion}\label{sec:discussion}
\subsection{Further Reducing Timesteps in BB Architectures}

\begin{table}[t]
\scriptsize
\centering

\setlength{\tabcolsep}{5pt}

\begin{tabular}{@{}c|c|c|c|c|c@{}}
\hline\hline
Factory & \makecell[c]{Syndrome\\Rounds} & $p_{\textrm{meas}}$ & $p_{\textrm{memory}}$ & $\tau_i$ & $p_{\textrm{out}}^{\textrm{(sim)}}$ \\   
\hline
$\textrm{8-to-CCZ}^{\otimes 2}_{\textrm{gross}}$ 
  & $7 \rightarrow 5$ 
  & $10^{-3.5}$ 
  & $10^{-5.2}$ 
  & $1570 \rightarrow [1159]$ 
  & $1.2 \times 10^{-4}$  \\
\hline
$\textrm{15-to-1}_{\textrm{gross}}$ 
  & $7 \rightarrow 4$ 
  & $10^{-2.7}$ 
  & $10^{-5.5}$ 
  & $6122 \rightarrow [3808]$ 
  & $3.5 \times 10^{-6}$  \\
\hline
$\textrm{20-to-4}_{\textrm{gross}}$ 
  & $7 \rightarrow 5$ 
  & $10^{-3.5}$ 
  & $10^{-5.2}$ 
  & $3088 \rightarrow [2581]$  
  & $7.8\times 10^{-5}$ \\
\hline\hline
\end{tabular}
\caption{Implementing magic-state distillation protocols in the gross code with fewer syndrome-extraction rounds. Entries of the form $x \rightarrow [y]$ denote the baseline timestep count and the reduced value at smaller $n$. This can speed up distillation with no or moderate impact on the output fidelity.}
\label{tab:improved}
\end{table}

In Section~\ref{sec:results}, we saw that BB-based distillation schemes achieve strong qubit savings at the cost of larger depth~$\tau_i$. Here, we explore a simple knob that reduces the number of syndrome-extraction rounds to lower $\tau_i$ in MSD circuits.

As observed in Ref.~\cite{cross2024improvedqldpcsurgerylogical}, the two error modes for in-module measurements, measurement-outcome flips $p_\textrm{meas}$ and logical memory errors $p_\textrm{memory}$, move in opposite directions as the number of rounds $n$ changes: $p_\textrm{meas}$ decreases roughly exponentially with $n$, while $p_\textrm{memory}$ grows approximately linearly \footnote{The syndrome extraction rounds and logical error rates are taken from the results of \cite{cross2024improvedqldpcsurgerylogical}, which is based on an LPU design that differs slightly from that in \cite{yoder2025tour} but yields comparable logical error rates.}.

General-purpose circuits typically choose $n$ near the balance point that minimizes total logical error. Distillation protocols, however, can tolerate some measurement flips, so their optimal $n$ can be smaller. In Section~\ref{sec:results}, we benchmarked how the parameter $\lambda$, the ratio between measurement error and the total logical operation error rate, affects the overall output error rate. Here, we reduce the number of syndrome-extraction rounds, thereby introducing more but tolerable outcome-flip errors while suppressing memory errors.

Table~\ref{tab:improved} reports the resulting estimated timestep reductions (shown in brackets). In the gross-code configuration, reducing $n$ from $7$ to $4$ for the 15-to-1 protocol cuts $\tau_i$ from $6122$ to $[3808]$ while keeping the output error at $p_{\textrm{out}}^{\textrm{(sim)}} \approx 3.5 \times 10^{-6}$. For 8-to-CCZ and 20-to-4, errors are already dominated by source errors, so decreasing $n$ slightly worsens fidelity but still yields substantial savings in $\tau_i$. In practice, operation-limited protocols can therefore trade a modest increase in $p_\textrm{meas}$ for noticeably shorter distillation time.

\subsection{Future Directions for Magic Factory Design and Fault-tolerant Computing in Bicycle Architecture}
Future improvements can be grouped into three directions. First, increasing the fidelity of in-module and inter-module logical operations through improved LPU design would directly lower output magic-state error and increase attainable throughput. Second, improving decoder performance can substantially improve logical operation fidelity and may provide a dominant improvement in practice. Third, it is important to study the impact of reducing the number of syndrome-extraction rounds: fewer rounds can lower latency and space-time cost, but may increase measurement-induced failures; quantifying this tradeoff would make the discussion above more actionable.

\section{Conclusion}
High-fidelity, resource-efficient magic-state distillation is critical for scalable fault-tolerant quantum computing. 
We introduced practical distillation factories on Bivariate Bicycle codes, which achieve low target error rates with competitive space–time costs while reducing the qubit footprint.
When preceded by surface code cultivation, our protocols constitute compelling two-round factories for near-term devices.
Looking ahead, our methodology is general and adaptive across protocols and hardware performance. As the bicycle architecture achieves lower logical error rates through better decoding, circuits, or devices, the factory design methodology in our paper delivers even lower output error rates with lower overhead, positioning BB-based magic state distillation as a robust building block for large-scale quantum platforms.

\section*{Acknowledgment}
We thank Steven M. Girvin, Andrew Cross, Theodore J Yoder, Tomas Jochym-O'Connor, Shraddha Singh, Zhixin Song, Xiang Fang, Ming Wang, Yue Wu, Shuwen Kan, Sean Garner, Samuel Stein, Chenxu Liu, and Ang Li for fruitful discussions. This work is supported in part by the National Science Foundation (under awards CCF-2312754 and CCF-2338063), in part by the U.S. Department of Energy, Office of Science, National Quantum Information Science Research Center, Co-design Center for Quantum Advantage (C2QA) under Contract No. DE-SC0012704, in part by QuantumCT (under NSF Engines award ITE-2302908), in part by AFOSR MURI (FA9550-26-1-B036). 
ZH acknowledges support from the MIT Department of Mathematics, the MIT-IBM Watson AI Lab, and the NSF Graduate Research Fellowship Program under Grant No. 2141064.
YD acknowledges partial support by Boehringer Ingelheim, and NSF NQVL-ERASE (under award OSI-2435244). External interest disclosure: YD is a scientific advisor to D-Wave Quantum, Inc.

\bibliographystyle{IEEEtran}
\bibliography{refs,sunny_refs}

\end{document}